\documentclass[aps, 10pt, prd, longbibliography,
               notitlepage, twocolumn, superscriptaddress,
               nofootinbib, floatfix]{revtex4-2}

\usepackage{amsmath}
\usepackage{placeins}
\usepackage[caption=false]{subfig}
\usepackage{amssymb}
\usepackage{amsfonts}
\usepackage[utf8]{inputenc}
\usepackage{comment}
\usepackage[T1]{fontenc}
\usepackage{mathrsfs}
\usepackage{anyfontsize}
\usepackage{microtype}

\usepackage[linktocpage,breaklinks]{hyperref}
\usepackage[dvipsnames]{xcolor}

\usepackage{tensor}
\usepackage{bm}

\usepackage{graphicx}
\graphicspath{{path/to/images/}}

\usepackage{epsfig}
\usepackage{epstopdf}

\usepackage{natbib}

\hypersetup{colorlinks=true,
            citecolor=NavyBlue,
            linkcolor=NavyBlue,
            urlcolor=NavyBlue}

\usepackage{tikz}

\newcommand{\dd}{{\rm d}}
\newcommand{\sr}{{\rm s}}

\newcommand{\ii}{{\rm i}}
\newcommand{\ee}{{\rm e}}
\newcommand{\lm}{\ell m}

\newcommand{\sw}{{}_{s}}

\newcommand{\DF}{Dudley--Finley\,}
\newcommand{\KN}{Kerr--Newman\,}
\newcommand{\BL}{Boyer--Lindquist\,}
\newcommand{\RN}{Reissner--Nordstr\"om\,}

\newcommand{\uiuc}{\affiliation{Department of Physics and Illinois Center for Advanced Studies of the Universe,\\
The Grainger College of Engineering,
University of Illinois Urbana-Champaign, Urbana, Illinois 61801, USA}}

\begin{document}
\title{Quasinormal modes of Kerr--Newman black holes:\\Revisiting the Dudley--Finley approximation}

\begin{abstract}
We present a comprehensive study of the \KN quasinormal mode spectrum in the
\DF approximation, where the linear gravitoelectromagnetic perturbations are
decoupled by ``freezing'' either one of the fields to its background value.
First, we reassess the accuracy of this approximation by comparing it to
calculations that solve the coupled system of gravitoelectromagnetic perturbation equations across the subextremal spin-charge parameter space.
We find that for the $(\ell,m,n) = (2,2,0)$, $(2,2,1)$, and $(3,3,0)$ modes,
the agreement is typically within $10\%$ and $1\%$ for the real and
imaginary parts of the frequencies, respectively.
Next, we investigate the spectrum in the near-extremal limit, and study the
family of long-lived (``zero-damped'') gravitational modes.
We find that the near-extremal parameter space consists of subregions
containing either only zero-damped modes, or zero-damped modes alongside modes
that retain nonzero damping. We derive analytic expressions for the boundaries
between these regions.
Moreover, we discuss the connection between the zero-damped and damped modes in
the \DF approximation and the ``near-horizon/photon-sphere'' modes of the full
\KN spectrum.
Finally, we analyze the behavior of the quadrupolar gravitational modes with large
overtone numbers $n$, and study their trajectories in the complex plane.
\end{abstract}

\author{Sagnik Saha}     \email{sagniks2@illinois.edu} \uiuc
\author{Hector O. Silva} \email{hosilva@illinois.edu}  \uiuc

\date{{\today}}

\maketitle

\section{Introduction} \label{sec:intro}

The \KN spacetime is the unique vacuum solution of the Einstein--Maxwell theory
in four spacetime dimensions representing a stationary, axisymmetric and
asymptotically-flat black hole~\cite{Newman:1965my}.
The solution is described by three parameters: the mass $M$, angular momentum
$J$, and electric charge $q$ of the black hole.
It reduces to the \RN solution when $J$ vanishes and to the Kerr solution when
$q$ vanishes. The Schwarzschild solution is recovered for vanishing $J$ and
$q$.
Given the uniqueness of the \KN solution~\cite{Mazur:1982db}, it is important to study its
stability.

The linear perturbations of the Einstein--Maxwell equations on a \KN spacetime
can be studied using the Newman--Penrose formalism~\cite{Newman:1961qr}.
The outcome is a system of coupled partial differential
equations describing the interaction between gravitational and electromagnetic
perturbations~\cite{Chandrasekhar:1978RSPSA.358..421C}.
Unlike what happens in the Kerr~\cite{Teukolsky:1972my,Teukolsky:1973ha},
\RN~\cite{Chandrasekhar:1979iz,Wald:1979}, and Schwarzschild~\cite{Bardeen:1973xb} cases, the latter two cases also including metric perturbations, these equations do not seem to be reducible to a system of
coupled ordinary differential equations, i.e., they are not separable in all
coordinates.
This nonseparability complicates both the stability analysis of the \KN solution and the characterization of its quasinormal modes, the latter being the focus of this work.

Quasinormal mode calculations of \KN black holes are often
done perturbatively, for example, in an expansion in small values of the
hole's spin~\cite{Pani:2013ija,Pani:2013wsa,Blazquez-Salcedo:2022eik} or
its charge-to-mass ratio~\cite{Mark:2014aja}.
Another approach consists of setting to zero (i.e., ``freezing'') either the
linear gravitational or electromagnetic perturbations of the \KN background.
This was the route taken by
Dudley and Finley~\cite{Dudley:1977zz,Dudley:1978vd}, who obtained a separable system of
equations that constitutes a deformation of the Teukolsky equation~\cite{Teukolsky:1972my,Teukolsky:1973ha}.
In this way, the quasinormal mode spectrum associated to the \DF equation
can be studied using standard techniques in black hole perturbation theory, such
as the Wentzel--Kramers--Brillouin (WKB) approximation~\cite{Kokkotas:1993ef}
or the continued fraction method~\cite{Berti:2005eb,Zimmerman:2015trm}; see also Ref.~\cite{Cano:2024jkd}.

In particular, Berti and Kokkotas~\cite{Berti:2005eb} compared the quasinormal
frequencies obtained from the \DF equation against the respective frequencies
obtained from a perturbed \RN black hole, where they froze either the
metric or the electromagnetic perturbations, thus analyzing the validity of the
\DF approximation.
Recently, Dias and
collaborators~\cite{Dias:2015wqa,Dias:2021yju,Carullo:2021oxn,Dias:2022oqm}
succeeded in calculating the quasinormal modes of \KN black holes directly from
the system of coupled partial differential equations, thus allowing for the first time a
systematic study of the mode spectrum across the spin-charge parameter space;
see Ref.~\cite{Berti:2025hly}, Sec.~2.2, for a review.
Henceforth, we refer to these calculations as
the ``full \KN solution,'' in the sense that no approximations, other than that the perturbations
are linear, are used.
The availability of these results invites for a reassessment of the accuracy of
the \DF approximation along the lines of Ref.~\cite{Berti:2005eb}. Carrying
this out is our first goal.

One outcome of Refs.~\cite{Dias:2021yju,Carullo:2021oxn} was the identification of different
branches of quasinormal modes, namely the near-horizon (NH) and the photon-sphere (PS)
modes.
This naturally raises the following question: what is the connection, if any, between these NH and PS
modes, and the damped and long-lived (``zero-damped'') modes that were identified
in earlier works in both Kerr~\cite{Hod:2008zz,Yang:2012pj,Yang:2013uba} (see also Ref.~\cite{Hod:2009td,Casals:2019vdb})
and \KN~\cite{Zimmerman:2015trm} spacetimes?
Answering this question is the second goal of this work.

Lastly, we survey, for the first time, quasinormal modes of the \KN solution in
the \DF approximation that are highly damped, i.e., modes with large
overtone numbers $n$. By doing so, we generalize earlier investigations in
the Kerr case~\cite{Berti:2003zu,Berti:2003jh}.

In the remainder of this paper we explain how we achieved these goals and what
we found. This work is organized as follows.
In Sec.~\ref{sec:DFeq}, we provide a concise review of the \DF equation,
followed by the calculation of its associated quasinormal frequencies in
Sec.~\ref{sec:numerical_methods}.
We revisit the accuracy of the \DF equation in
Sec.~\ref{sec:accuracy_df_vs_kn}.
Then, in Sec.~\ref{sec:zdm_dm_near_extremality}, we study the gravitational zero-damped modes across the spin-charge parameter space.
We then draw some connections between the NH and PS modes of the full \KN solution and the damped and zero-damped modes of the \DF approximation in Sec.~\ref{sec:nh_ps_zdm_dm_connection}.
Finally, in Sec.~\ref{sec:highly_damped_df_qnms}, we investigate the quadrupolar
gravitational quasinormal modes at large overtone numbers and discuss some of
their properties.
We summarize our conclusions in Sec.~\ref{sec:conclusions}.

We use dimensionless geometrical units $c = G = 2 M = 1$, and the mostly plus metric signature.

\section{The \DF equation} \label{sec:DFeq}

As we described in Sec.~\ref{sec:intro}, the coupled gravitoelectromagnetic
perturbations of the \KN solution are not separable when decomposed in modes;
see Ref.~\cite{Chandrasekhar:1985kt}, Sec.~111, or Ref.~\cite{Giorgi:2020ujd}
for a discussion.
In Refs.~\cite{Dudley:1977zz,Dudley:1978vd}, Dudley and Finley studied the
separability of the linear perturbations of the
Pleba\'{n}ski--Damia\'{n}ski~\cite{Plebanski:1976gy} family of solutions of the
Einstein--Maxwell theory.
This family of solutions encompasses all electrovacuum spacetimes of
Petrov-type D, including the \KN solution~\cite{Newman:1965my}.
Dudley and Finley found that mode-separability of the perturbations is possible
if either the background metric or the Maxwell field is kept fixed.
This approximation results in a differential equation for the radial dependence
of the perturbations whose functional form is similar to that of the Teukolsky
equation~\cite{Teukolsky:1972my,Teukolsky:1973ha}, and is now known as the \DF
equation.
The angular dependence of the perturbations is described by the spin-weighted
spheroidal harmonics~\cite{Teukolsky:1972my,Teukolsky:1973ha}.

Expressed in \BL coordinates, the \DF equation reads
\begin{align}
\label{eq:DFeq}
\begin{aligned}
&\Delta^{-s} \frac{\dd}{\dd r} \left[ \Delta^{s+1} \frac{\dd \, \sw R_{\lm}}{\dd r} \right]
+ \frac{1}{\Delta} \left[ K^2 - \ii s \frac{\dd \Delta}{\dd r} K \right.
\\
&\quad
\left.
+ \, \Delta \left( 2 \ii s \frac{\dd K}{\dd r} - \sw \lambda_{\lm} \right)
\right] \, \sw R_{\lm} = 0,
\end{aligned}
\end{align}
where $\omega$ is the frequency of the mode $\sw R_{\lm}$,
and where we defined
\begin{subequations} \label{eq:def_K_and_D}
    \begin{align}
        K = (r^2 + a^2) \, \omega - a m,
        \\
        \Delta = (r - r_{+})\,(r - r_{-}),
    \end{align}
\end{subequations}
and
\begin{equation}
    r_\pm = (1 \pm b) / 2 \,, \quad \textrm{where} \quad b = \sqrt{1 - 4 (a^2 + q^2)},
\label{eq:rpm}
\end{equation}
are the locations of the outer, $r_{+}$, and inner, $r_{-}$, horizons in \BL coordinates.
The black hole's angular momentum per unit mass, $a$, is bound to the interval
\begin{equation}
    0 \leq a < \sqrt{1 - 4 q^2} /  2\,,
\end{equation}
in our units, $2 M  = 1$. In addition,
\begin{equation} \label{eq:def_slambdalm}
    \sw\lambda_{\lm} = \sw A_{\lm} + (a \omega)^2 - 2 am \omega,
\end{equation}
where $\sw A_{\lm}$ is a separation constant determined as the eigenvalue of
the spin-weighted spheroidal harmonic equation;
see~Eq.~\eqref{eq:S_harmonic_eq}.
The spin-weight parameter $s$ has values $0$, $-1$, and $-2$ for scalar,
electromagnetic, and gravitational perturbations, respectively.
We will focus on the gravitational case in our numerical calculations.

The \DF equation is exact only in two cases. First, when $s = 0$, it
describes the perturbations of a massless scalar field to the \KN
solution. Second, when $q = 0$, it reduces to the Teukolsky equation~\cite{Teukolsky:1972my,Teukolsky:1973ha}.
Consequently, for $q = 0$, it further reduces to the Bardeen--Press equation when the spin $a$ vanishes~\cite{Bardeen:1973xb}.
A variation of the \DF equation that has the same quasinormal mode spectrum as
Eq.~\eqref{eq:DFeq} was presented in Ref.~\cite{Silva:2025khf} that followed
the ideas of Refs.~\cite{Casals:2021ugr,Aminov:2020yma,Hatsuda:2020iql} in its derivation.

In general, when $a$ is nonzero, the separation constant $\sw A_{\lm}$ is
determined, for a given value of $\omega$, by solving the spin-weighted
spheroidal harmonic equation
\begin{align} \label{eq:S_harmonic_eq}
    &\frac{\dd}{\dd u} \left[ (1 - u^2) \frac{\dd \, \sw S_{\lm}}{\dd u} \right]
    + \biggl[ (a \omega u)^2 - 2 a \omega s u + s + \sw A_{\lm} \phantom{\frac{}{}}  \nonumber \\
    &\qquad - \frac{(m + su)^2}{1-u^2} \biggr] \, \sw S_{\lm} = 0,
\end{align}
where $u = \cos \vartheta$ is related to the polar angle $\vartheta$ of the \BL coordinates.
We impose boundary conditions such that the eigenfunctions $\sw S_{\lm}$ are
finite at the regular points $u = \pm 1$.
These eigenfunctions are known as the spin-weighted spheroidal harmonics, and
$c = a\omega$ is the spheroidicity parameter. This parameter is complex valued
in general.
For vanishing spheroidicity, the eigenfunctions $\sw S_{\lm}$ become the
spin-weighted spherical harmonics, with eigenvalues
\begin{equation}
    \sw A_{\lm} = \ell ( \ell + 1) - s (s + 1)\,,
    \quad \textrm{for} \quad c = 0.
\end{equation}
Corrections to this expression for small values of $c$ can be obtained
perturbatively; see, for instance,
Refs.~\cite{Press:1973zz,Fackerell:1977,Seidel:1988ue,Berti:2005gp}.

\section{Numerical methods} \label{sec:numerical_methods}

In this section, we present an overview of the numerical techniques that
we used to compute the quasinormal frequencies of the Dudley-Finley equation~\eqref{eq:DFeq},
along with a summary of the ways in which we validated our codes.

\subsection{Calculation of quasinormal modes using continued fractions} \label{sec:leaver}

We use Leaver's method to compute the quasinormal frequencies~\cite{Leaver:1985ax}.
This method was first applied to the \DF equation in Ref.~\cite{Berti:2005eb}.
We present a synopsis of this method to make our work fairly self-contained.

The starting point consists in observing that, by requiring the functions $\sw R_{\lm}$
to be purely ingoing at the (outer) event horizon $r_+$ and purely outgoing at spatial infinity
(the so-called ``quasinormal-mode boundary conditions''), $\sw R_{\lm}$ behaves as
\begin{align}
\begin{aligned}
\label{eq:qnm_bcs}
    \lim_{r\,\to\,r_+}    \sw R_{\lm} &\simeq \, (r - r_+)^{-s - \ii \sigma_+},
    \\
    \lim_{r\,\to\,\infty} \sw R_{\lm} &\simeq \, r^{-1-2s+\ii \omega} \, \ee^{\ii \omega r},
\end{aligned}
\end{align}
where
\begin{equation}
\sigma_{+} = [\omega (r_+ - q^2) - am]/b,
\end{equation}
%
and the inner and outer horizon locations $r_\pm$ and $b$ are given by Eq.~\eqref{eq:rpm}.
A solution to Eq.~\eqref{eq:DFeq} satisfying the boundary conditions~\eqref{eq:qnm_bcs} can be written
in the form~\cite{Leaver:1985ax}
\begin{align}
    \sw R_{\lm} &= \ee^{\ii \omega r}
    \, (r - r_-)^{-1-s+\ii\omega+\ii\sigma_+}
    \, (r - r_+)^{-s-\ii\sigma_+}
    \nonumber \\
                &\quad \times \sum_{n=0}^{\infty} a_{n} \left(\frac{r-r_+}{r-r_-}\right)^{n} \,.
    \label{eq:leaver_df}
\end{align}
Substituting Eq.~\eqref{eq:leaver_df} into Eq.~\eqref{eq:DFeq} yields a three-term recursion relation for the coefficients $a_n$, that we write as:
\begin{align}
\begin{aligned}
    &\alpha_{0} \, a_{1} + \beta_{0} \, a_0 = 0\,, \\
    &\alpha_{n} \, a_{n+1} + \beta_{n} \, a_{n} + \gamma_{n} \, a_{n-1} = 0\,,
    \quad n = 1,\,2,\, \dots
\end{aligned}
\end{align}
The coefficients in the recursion relation are:
\begin{align} \label{eq:coef_abc_radial}
\begin{aligned}
    \alpha_{n} &= n^2 + (c_0 + 1) n + c_0 \,,             \\
    \beta_{n}  &= -2 n^2 + (c_1 + 2) n + c_3\,,           \\
    \gamma_{n} &= n^2 + (c_2 - 3) n + c_4 - c_2 + 2\,.
\end{aligned}
\end{align}
The additional coefficients $c_i$ ($i = 0,\dots,4$) encode information about
the black hole's spin $a$ and charge $q$, the spin-weight $s$ of the perturbing field, the azimuthal index $m$ of the perturbation, and the angular separation
constant $\sw A_{\ell m}$.
The explicit expressions of $c_i$ can be found in Ref.~\cite{Berti:2005eb},
Eq.~(6), or in Ref.~\cite{Silva:2025khf}, Eq.~(30).

The series~\eqref{eq:leaver_df} converges and the boundary
conditions~\eqref{eq:qnm_bcs} are thus satisfied if $\omega$ is a solution of
the continued fraction
\begin{equation} \label{eq:cf_rad}
    0 = \beta_0
    - \frac{\alpha_0 \, \gamma_1}{\beta_1 -}
    \,\frac{\alpha_1 \, \gamma_2}{\beta_2 -}
    \,\frac{\alpha_2 \, \gamma_3}{\beta_3 -} \cdots,
\end{equation}
for given values of $a$, $q$, $s$, $m$, and $\sw A_{\ell m}$. The angular
separation constant $\sw A_{\ell m}$ is obtained by solving a similar
continued-fraction equation associated to Eq.~\eqref{eq:S_harmonic_eq}; see
Ref.~\cite{Leaver:1985ax}, Eqs.~(20) and (21). Its derivation follows the same
steps outlined for the radial equation.

To obtain a quasinormal frequency, the radial and angular continued fractions
must be satisfied simultaneously. Hence, the problem of obtaining a quasinormal
frequency reduces to a double root-finding problem.
In practice, we sum the continued fractions from bottom to top, starting from a
large truncation index $N$ of the order of $10^3$.

Leaver~\cite{Leaver:1985ax} observed that the $n$th overtone is numerically
the most stable root of the $n$th inversion of the radial continued
fraction; see Eq.~(14) therein.
The fundamental frequency is obtained using the noninverted form~\eqref{eq:cf_rad}.
When computing overtones, we follow Nollert~\cite{Nollert:1993zz}, who showed that the convergence
of the summation of the continued fraction from ``bottom to top'' is improved
if the remainder of the continued fraction,
\begin{equation}
R_N = \frac{\gamma_{N+1}}{\beta_{N+1}-\alpha_{N+1}\,R_{N+1}}\,,
\end{equation}
is approximated by a power series in $N^{-1/2}$,
\begin{equation}
R_N = \sum_{k=0}^{\infty}C_k \, N^{-k/2}.
\end{equation}
The coefficients $C_k$ for the radial \DF equation up to $k = 2$ can be found
in Ref.~\cite{Berti:2005eb}.

To locate the roots, we used Muller's method~\cite{Muller:MR0083822},
following the pseudocode from Chapter~9.2 in Ref.~\cite{Press2007Numerical}.
%
We also used a perturbative expansion in powers of $c$ for the separation
constant $\sw A_{\lm}$ from Ref.~\cite{Berti:2005gp} to set our initial guesses
for the root-finding procedure applied to the angular continued fraction.

For our numerical calculations, we used two codes written independently in
Mathematica and C++, both of which we validated by reproducing the results of
Ref.~\cite{Berti:2005eb} for the \DF equation, and those of Ref.~\cite{BertiTab}
in the Kerr and Schwarzschild limits.

\section{The accuracy of the \DF approximation} \label{sec:accuracy_df_vs_kn}

Although the \DF equation is expected to describe the quasinormal spectrum of the \KN solution only \emph{qualitatively}, it is of intrinsic interest to know
\emph{quantitatively} how accurate the approximation is.
As we described in Sec.~\ref{sec:intro}, such a quantitative comparison was done
in the nonrotating limit (i.e., by comparing against the quasinormal
spectrum of the \RN solution) in Ref.~\cite{Berti:2005eb}. Later, it was also
carried out for slowly-rotating \KN black holes in Ref.~\cite{Pani:2013wsa}.
A detailed study of the quasinormal spectrum of the \KN solution spanning the
full spin-charge parameter space was only possible (and executed) later by
Dias~et~al.~\cite{Dias:2015wqa,Dias:2021yju,Carullo:2021oxn,Dias:2022oqm}.
The goal of this section is to reassess the validity of the \DF equation in light of
these developments, thereby updating the analyses of Refs.~\cite{Berti:2005eb,Pani:2013wsa}.
To carry out this comparison, we use two sets of results from these works.
\begingroup
\allowdisplaybreaks
\begin{enumerate}
    \item \emph{Numerical results:} These were obtained in Ref.~\cite{Dias:2015wqa} and
        are publicly available at~\cite{BertiTab}.
        This data corresponds to the $(\ell,m,n) = (2,2,0)$ mode computed along the line $a = q$.
        The Schwarzschild and extremal limits correspond to $a = q = 0$ and $a = q = (2\sqrt{2})^{-1} \approx 0.35$, respectively.
    \item \emph{Bayesian fitting formulas:} These were presented in
        Ref.~\cite{Carullo:2021oxn}, and provide accurate fits for
        the $(\ell,m,n) = (2,2,0)$, $(2,2,1)$, and $(3,3,0)$ modes computed over
        the spin-charge parameter space.
\end{enumerate}
\endgroup

We quantify the accuracy of the \DF equation through the logarithmic absolute error:
\begin{equation} \label{eq:def_log_abs_diff}
    \Delta \omega_{\ell m n} = \log_{10}|\omega^{\rm DF}_{\ell m n} - \omega^{\rm KN}_{\ell m n}|,
\end{equation}
between \emph{gravitational, $s=-2$,} quasinormal frequencies obtained from the \DF equation, $\omega^{\rm DF}_{\ell m n}$,
and by solving the complete system of partial differential equations that occur in the \KN problem,
$\omega^{\rm KN}_{\ell m n}$. The latter can represent either the ``raw'' numerical results or the results from the fitting formulas.
We use Eq.~\eqref{eq:def_log_abs_diff} as a shorthand notation for the absolute error
between the real and imaginary parts of the two frequencies.
We focus on the gravitational perturbations for this comparison.

\subsection{Comparison along $a = q$}

\begin{figure}[t]
\includegraphics{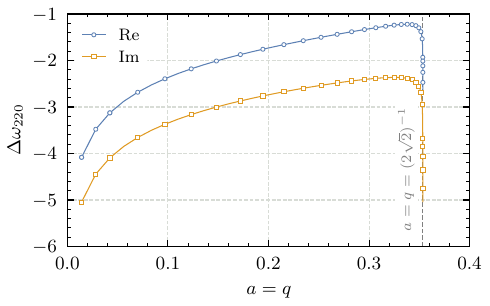}
\caption{Logarithmic absolute error between the $(\ell, m, n) = (2, 2, 0)$ quasinormal frequencies calculated in the \KN problem and with the \DF equation for $s=-2$, for points on the line $a = q$ in the spin-charge parameter space. We show the errors in the real (``Re'') and imaginary (``Im'') parts of the frequencies. The dashed vertical line at $a = q = (2 \sqrt{2})^{-1}$ indicates the extremal limit.}
\label{fig:l2m2n0exactvsdf}
\end{figure}

We begin by comparing the Dudley-Finley approximation to the numerical results from
Ref.~\cite{Dias:2015wqa}. 
In Fig.~\ref{fig:l2m2n0exactvsdf}, we show $\Delta \omega_{220}$ as we move along the line $a = q$. The two curves represent absolute errors in the real (``Re'') and imaginary
(``Im'') parts of the frequencies, as defined in
Eq.~\eqref{eq:def_log_abs_diff}.
We omit the Schwarzschild limit in this figure because, as expected, the agreement between
the two calculations is excellent, of the order $10^{-12}$.
As we move further along this line, the error increases monotonically until it reaches a maximum value of approximately $10^{-1}$ for the real part of the frequency, at a point
close to extremality, $a = q = 0.35285$.
Surprisingly, beyond this point, the errors for both the real and imaginary parts of
the frequency decrease sharply again, reaching a minimum value of
approximately $10^{-5}$ near extremality.
Interestingly, we find that the error in the imaginary part of the frequency
remains below $1\%$ everywhere.

\begin{figure}
    \includegraphics{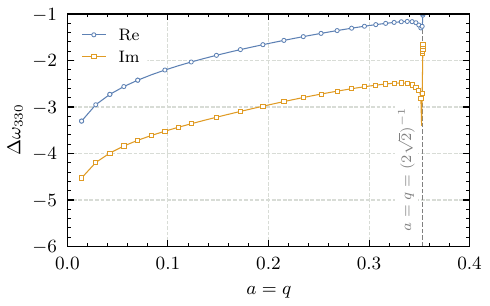} \\
    \includegraphics{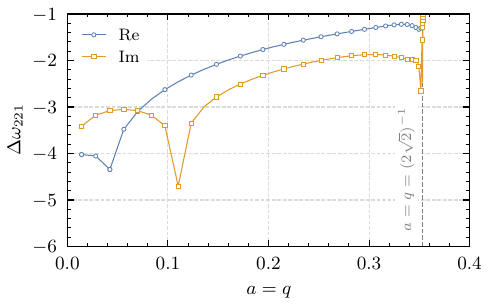}
    \caption{Logarithmic absolute error between the quasinormal frequencies calculated using the Bayesian fitting formula and the Dudley–Finley approximation for $s=-2$, for points on the line $a = q$ in the spin-charge parameter space. We show the results for two sets of $(\ell, m, n)$ values: $(3,3,0)$ in the top panel and $(2,2,1)$ in the bottom panel. As in Fig.~\ref{fig:l2m2n0exactvsdf}, we show the errors in the real (``Re'') and imaginary (``Im'') parts of the frequencies. The dashed vertical line at $a = q = (2 \sqrt{2})^{-1}$ indicates the extremal limit.}
    \label{fig:knfit_vs_df_aeq}
\end{figure}

Are these results shared by the other modes? To answer this question, we now use the fitting formulas from Ref.~\cite{Carullo:2021oxn} to carry out the same comparison as for the $(2,2,0)$ mode, this time focusing on the $(2,2,1)$ and $(3,3,0)$ modes.
The outcome of this exercise is summarized in Fig.~\ref{fig:knfit_vs_df_aeq}.
As before, we omit the Schwarzschild limit, since the two methods agree very well there.
For the $(3,3,0)$ mode (top panel), the agreement is initially excellent but gradually deteriorates, reaching a maximum error at $a = q = 0.33801$.
Beyond this point, the error momentarily decreases before rising sharply again.
Apart from this, the behavior of $\Delta\omega_{330}$ is qualitatively
the same as that of $\Delta\omega_{220}$; cf.~Fig.~\ref{fig:l2m2n0exactvsdf}.
For the $(2,2,1)$ mode (bottom panel), the behavior is more interesting:
$\Delta\omega_{221}$ initially increases along the $a = q$ line, then undergoes sharp drops away from extremality, first in the real part, and then in the imaginary part.
Beyond these points, $\Delta \omega_{221}$ increases until a critical point near extremality, after which, similar to $\Delta \omega_{330}$, it momentarily decreases before increasing sharply again.

\begin{figure}
\includegraphics[width=\columnwidth]{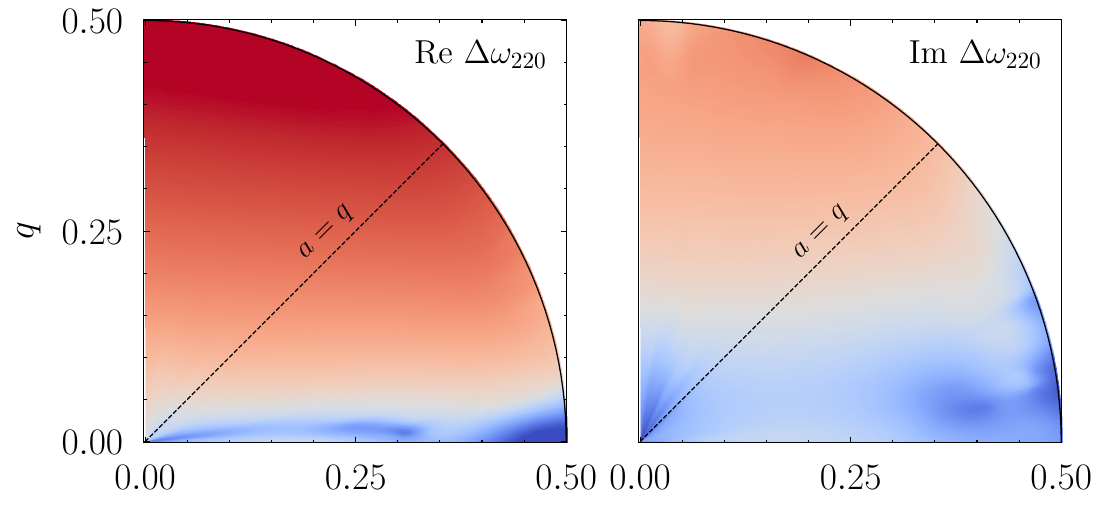} \\
\includegraphics[width=\columnwidth]{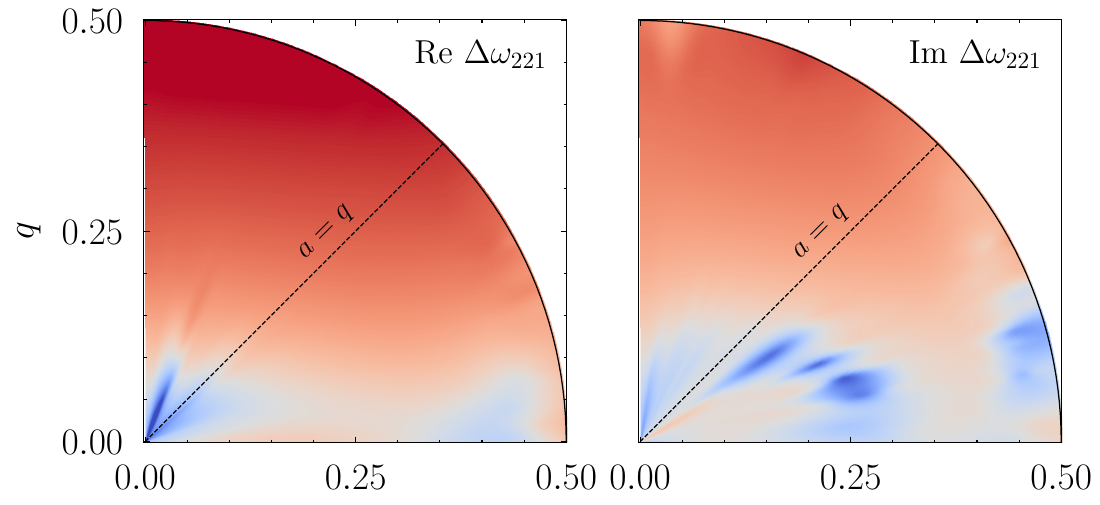} \\
\includegraphics[width=\columnwidth]{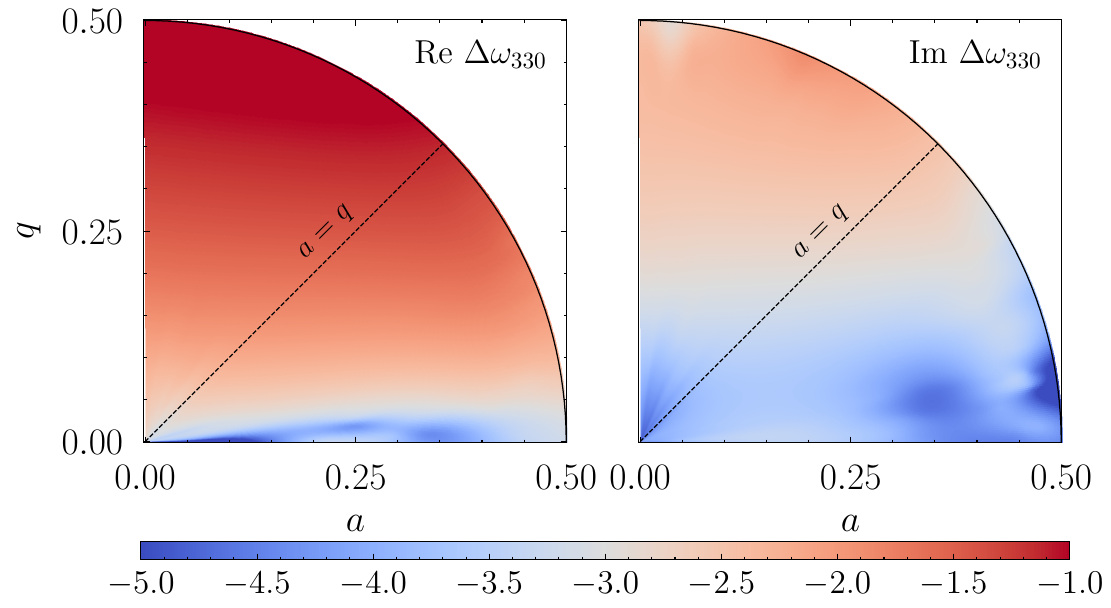}
\caption{Logarithmic absolute error between the quasinormal frequencies computed using the Bayesian fitting formula and the \DF approximation for $s = -2$, for points within the quarter-circular region defined by $a^2 + q^2 \leq 1/4$. We show the results for $(\ell,m,n) = (2,2,0), (2,2,1)$, and $(3,3,0)$ as heat maps.
The dashed line corresponds to the $a = q$ line that we studied in Figs.~\ref{fig:l2m2n0exactvsdf} and~\ref{fig:knfit_vs_df_aeq}. In the near-extremal region, the errors increase as we move from the high-spin to the high-charge limit. The \DF approximation yields about $1\%$ errors for the imaginary parts of all the mode frequencies that we studied for subextremal \KN black holes. Moreover, for small values of $q$, the errors in the real parts of these frequencies are also restricted to about $1\%$.}
\label{fig:knfit_vs_df_circ}
\end{figure}

In summary, we found that for the $(\ell, m, n) = (2,2,0)$, $(2,2,1)$, and
$(3,3,0)$ modes, calculated along the line $a = q$ from the Schwarzschild limit
up to the near-extremal limit, the \DF equation reproduces the \KN quasinormal
frequencies with percent absolute errors below 10\% and 1\% in the real and
imaginary parts, respectively.
For the $(2,2,1)$ and $(3,3,0)$ modes we also calculated the absolute relative
errors $|1 - \omega^{\rm DF}_{\ell m n}/\omega^{\rm KN}_{\ell m n}|$.
We found that the percent mean absolute relative error between the Bayesian fits
and the \DF approximation is approximately $3\%$ for
the real part and $13\%$ for the imaginary part of the
frequencies. These values are much larger than the percent relative error between the Bayesian
fits and the numerical data used to obtain the fits, which is typically of the order
$\pm 0.2\%$~\cite{Carullo:2021oxn}.
Hence, we attribute the errors for all modes, as seen in Figs.~\ref{fig:l2m2n0exactvsdf}
and~\ref{fig:knfit_vs_df_aeq}, to our use of the \DF approximation.

\subsection{Comparison over the wider parameter space}

Next, we perform a similar comparison for the same set of $(\ell, m, n)$ values,
but now in the quarter circular region in parameter space defined as follows:
\begin{equation}
    a^2 + q^2 \leq 1/4, \hspace{2mm} 0 \leq (a,q) \leq 1/2.
    \label{eqref:extremalkncondition}
\end{equation}
(Recall that we use units in which $2M=1$.)
The boundary of this region corresponds to an extremal black hole, beyond which
the \KN solution represents a naked singularity.
To evaluate $\omega^{\rm KN}_{\ell mn}$ in the
region~\eqref{eqref:extremalkncondition}, we use the Bayesian fitting formulas of Ref.~\cite{Carullo:2021oxn}.

As before, we use Eq.~\eqref{eq:def_log_abs_diff} to quantify our comparison.
However, now we evaluate it on 300 points uniformly distributed within
the region defined in Eq.~\eqref{eqref:extremalkncondition}, followed by an
interpolation onto a $300 \times 300$ grid. For this, we use the radial basis
function method with a ``thin plate spline'' kernel; see Chapter $3.7.1$ in
Ref.~\cite{Press2007Numerical}. We verified that the interpolation did not introduce any spurious features that were not present in the original grid data. Figure~\ref{fig:knfit_vs_df_circ} shows our results in the form of heat maps.

In the near-extremal regime, the errors are much larger in the high-charge
limit than in the high-rotation limit.
This likely reflects the breakdown of the \DF approximation at large values of
the black hole charge $q$, where neglecting the coupling between
electromagnetic and gravitational perturbations is
unjustified~\cite{Berti:2005eb}. Also, this may reflect the omission of the NH
modes of the full \KN problem from the fitting procedure, as these modes
dominate the spectrum in the near-extremal, high-charge
limit~\cite{Carullo:2021oxn}.
We postpone a detailed discussion of the families of modes in the full \KN problem to Sec.~\ref{sec:nh_ps_zdm_dm_connection}.

In the subextremal region, the absolute errors in the imaginary parts of the frequencies are generally limited to approximately $1\%$, while they can be slightly larger for the real parts. However, when the black hole's charge $q$ is small, the absolute errors in \textit{both} the real and imaginary parts remain below 1\%, independent of $a$.

\section{Zero-damped and damped modes near extremality} \label{sec:zdm_dm_near_extremality}

We now study the quasinormal mode spectrum of the \KN spacetime near
extremality using the \DF equation.
Our investigation is guided by earlier studies of the near-extremal Kerr
spacetime, which showed that the quasinormal mode spectrum in this regime
exhibits rich and intricate features, particularly for corotating modes with $m
\geq 0$~\cite{Hod:2008zz,Yang:2012pj,Yang:2013uba}.
For these modes, the spectrum splits above a certain value of the spin $a$ into
two distinct branches, the DMs and ZDMs, characterized by their decay rates.
Specifically, the ZDMs are characterized by much slower decay rates than the DMs.
The situation for the near-extremal \KN solution turns out to be similar, although with additional features from the interplay between the charge and angular momentum of the black hole.

In this section, to lighten the notation, we will omit the subscripts $\ell$,
$m$, $n$, and $s$ from the various quantities.

\subsection{Expression for the zero-damped modes near extremality}

We study the properties of the ZDMs using tools adapted from the Kerr spacetime with suitable modifications. The first of these tools is the technique of
matched asymptotic expansion which, when applied to the \DF equation, furnishes
the following expression for the \KN ZDM
frequencies~\cite{Zimmerman:2015trm},
\begin{equation}
    \omega = m \Omega_H |_{\rm ext} - \frac{\sigma}{1 + 4 a^2}\bigg[\delta + \ii \bigg(n + \frac{1}{2} \bigg) \bigg] + \mathcal{O}(\sigma^2),
\label{eqref:zdmeqn}
\end{equation}
where the angular frequency of the horizon, $\Omega_H$, in the extremal limit, $r_+ = M$, is,
\begin{equation} \label{eq:omega_H_extremal}
\Omega_{H}|_{\rm ext} =
\left. \frac{a}{r_+^2 + a^2} \right\vert_{r_+ = M}
= \frac{a}{2a^2 + q^2}
= \frac{4a}{1 + 4 a^2}.
\end{equation}
In going from the second to the third equality, we used the fact that at extremality, $q^2 = 1/4 - a^2$.
The other quantities in Eq.~\eqref{eqref:zdmeqn} are defined as,
\begin{subequations}\label{eqref:wzdmparamdef}
\begin{align}
\sigma       &= 1 - {r_-}/{r_+}, \quad \text{(``off-extremal parameter'')}\\
\delta^2     &= (2\omega r_+)^2 - (s + 1/2)^2 - \lambda,
\end{align}
\end{subequations}
and $\lambda$ was defined in Eq.~\eqref{eq:def_slambdalm}.

Qualitatively, the derivation of Eq.~\eqref{eqref:zdmeqn} involves solving Eq.~\eqref{eq:DFeq} in two distinct regions—an outer, far-field region and an inner, near-horizon region—followed by a matching procedure in an intermediate region where both approximations are valid. See Ref.~\cite{Zimmerman:2015trm} for details. The derivation further assumes that,
\begin{equation} \label{eq:zdm_asumpt}
\omega - m\Omega_H \ll 1,
\end{equation}
implying that only those modes with $m\geq0$ (the co-rotating modes) contain ZDMs in their spectrum (since $\rm Re \ \omega > 0$).
We emphasize that Eq.~\eqref{eqref:zdmeqn} is valid to leading order in the off-extremal parameter
$\sigma$.

\subsection{WKB criteria for the coexistence of zero-damped and damped modes} \label{sec:wkb_criteria}

Next, we perform a WKB analysis of the perturbation Eqs.~\eqref{eq:DFeq} and~\eqref{eq:S_harmonic_eq}, to obtain an analytic condition defining the boundary between subregions (in the spin-charge parameter space) containing only ZDMs and those containing both ZDMs and DMs.
This result is remarkably accurate even for $\ell = 2$, despite being formally valid only for $\ell \gg 1$ (eikonal limit).
Here we summarize the key ideas and present the final result; further details can be
found in Refs.~\cite{Yang:2012pj,Yang:2013uba,Zimmerman:2015trm}. Defining
\begin{equation}
L = \ell + 1/2
\quad \textrm{and} \quad
\mu = m/L,
\end{equation}
we first expand the radial~\eqref{eq:DFeq} and angular~\eqref{eq:S_harmonic_eq} equations to leading in order in $L$. As a result, the radial equation takes the following form:
\begin{equation}
    \left[\frac{\dd^2}{\dd r_*^2} - V_r \right] u = 0,
    \quad
    u(r) = \Delta^{s/2}\sqrt{r^2 + a^2} R(r),
    \label{eq:DFeq_WKB}
\end{equation}
where $V_r$ denotes the leading order WKB potential,
\begin{equation}
    V_r = -\frac{K^2 - \lambda \,\Delta}{(r^2 + a^2)^2}.
    \label{eq:wkb_vr}
\end{equation}
and $K$ and $\Delta$ are defined in Eq.~\eqref{eq:def_K_and_D}.
Equation~\eqref{eq:DFeq_WKB} is a familiar Schr\"odinger-like equation, with $V_r$ representing the effective potential. This allows us to perform a WKB expansion around the peak of the potential, $r_p$. In the extremal limit, an examination of the location of the peak $r_p$ in relation to the horizon $r_+$ yields the analytic condition that defines the boundary between the two regimes.
Specifically, if $r_p$ lies on the horizon, only ZDMs are present; if additional peaks exist \textit{outside} the horizon, both DMs and ZDMs are supported.
Mathematically, these arguments lead to the conditions:
\begin{equation} \label{eq:summary_mu_criteria}
   \begin{split}
       &\mu < \mu_c \implies \text{ZDMs and DMs coexist.}\\
       &\mu \geq \mu_c \implies \text{Only ZDMs exist,}\\
   \end{split}
\end{equation}
where the critical value $\mu_c$ is defined as:
\begin{equation} \label{eq:def_muc}
    \mu_c^2 = \frac{1}{2}\left[3 + \frac{12 - \sqrt{136 + 224a^2 + 16a^4}}{4a^2}\right].
\end{equation}
Here, we used the extremality condition $q^2 = 1/4 - a^2$ to eliminate the
explicit dependence on the charge $q$ in this expression. As such,
Eq.~\eqref{eq:def_muc} is strictly valid at extremality only, although we will
occasionally evaluate it in the near-extremal region.
Because $\mu$ is at most one, we see that if $a < a_c = 1/4$, where $a_c$ is the critical spin value obtained by setting $\mu_c^2 = 1$, the coexistence of DMs and ZDMs \emph{is guaranteed}, since the condition $\mu < \mu_c$ is necessarily satisfied in such a case.

\subsection{General criteria for the coexistence of zero-damped and damped modes} \label{sec:dsqfsq}

Motivated by the WKB analysis, we now introduce an alternative analytic criteria
to identify the boundary between the subregions containing only ZDMs and those
containing both ZDMs and DMS that is valid for \emph{general} $\ell$.
While this criteria has already been established for the Kerr case~\cite{Yang:2013uba} (see also Ref.~\cite{Detweiler:1977gy}) to our knowledge, this is the first time that it has been extended to the \KN case in the \DF approximation.

The key insight is that although the role of the parameter $\mu$ (and hence of $\mu_c$) is unclear for general $\ell$, the potential peak is still expected to play the same role as before. With this in mind, we proceed by expressing the radial \DF equation~\eqref{eq:DFeq} in the extremal limit, assuming frequencies of the form $\omega = m\Omega_H + \mathcal{O}(\sigma)$.
For scalar perturbations ($s = 0$), this procedure immediately yields an expression of the same form as Eq.~\eqref{eq:DFeq_WKB}, with a real-valued potential $V_0$. For electromagnetic ($s = -1$) and gravitational ($s = -2$) perturbations, however, we initially obtain \textit{complex-valued} potentials. Thankfully, by suitable transformations, we can still obtain real-valued potentials in both cases. Details of these transformations can be found in Refs.~\cite{Detweiler:1977gy,Yang:2013uba} for the Kerr spacetime, which we extended to the \KN case.
We find:
\begin{equation}
    \left[\frac{\dd^2}{\dd r_*^2} - V_s\right]R_s = 0,
    \quad s = \{0,-1,-2\}.
    \label{eq:scat}
\end{equation}
where $R_s$ and $V_s$ denote the suitably transformed radial functions and potentials, respectively. We present the explicit forms of the potentials $V_s$ in Appendix~\ref{app:transformed_potentials}.
We can then show that the following conditions hold at $r_+$:
\begin{equation} \label{eq:vs_hor_conds}
    V_s(r_+) = 0, \quad\textrm{and}\quad V_s'(r_+) = 0.
\end{equation}
They imply that there necessarily exists a local extremum of the potential at the horizon.
However, for this to be the \textit{only peak} (globally), the additional condition
\begin{equation} \label{eq:d2V_dr2}
V_{s}''(r_+) < 0
\end{equation}
must also be satisfied. Equation \eqref{eq:d2V_dr2} is not a trivial statement; we discuss it further in Appendix~\ref{app:potential_extrema}. For this additional condition to hold, an equivalent
function, $\mathcal{F}_s^2$, must be positive. In summary:
\begin{align}
    \mathcal{F}_s^2 > 0 \implies \text{peak only at $r_+$} \implies \text{only ZDMs},
    \label{eq:nonwkb_zdmcond}
\end{align}
where $\mathcal{F}_s^2$ for $s=0$, $-1$, and $-2$ is given as follows:
\begin{subequations}\label{eq:fsq_defn_kn}
\begin{align}
    \mathcal{F}_0^2 &= \frac{16a^4m^2 + 24 a^2m^2}{(4a^2 + 1)^2} - A, \\
    \mathcal{F}_{-1}^2 &= \frac{16a^4(m^2 - 1) + 8 a^2(1+3m^2) - 1}{(4a^2 + 1)^2} - A, \\
    \mathcal{F}_{-2}^2 &= \frac{16a^4(m^2 - 1) + 8 a^2(3m^2 - 1) - 1}{(4a^2 + 1)^2} - A.
\end{align}
\end{subequations}
Above, $A$ is the separation constant, cf.~Eq.~\eqref{eq:def_slambdalm},
evaluated at $\omega = m\Omega_H |_{\rm ext}$.
The dependence on the charge $q$ in these expressions is implicitly contained
in the horizon angular frequency $\Omega_H$ at extremality, cf.~Eq.~\eqref{eq:omega_H_extremal}.

\begin{figure}[t]
    \includegraphics{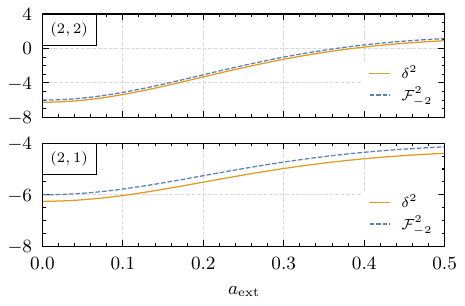}
    \caption{Comparison between $\delta^2$ and $\mathcal{F}_s^2$ for the fundamental $(\ell,m) = (2,2)$ and $(2,1)$ gravitational modes (top and bottom panel, respectively). We show the behavior of $\delta^2$ and $\mathcal{F}_{s}^2$ as we move along the extremal curve $a_{\rm ext}^2 + q_{\rm ext}^2 = 1/4$. In general, the two curves follow each other closely. In the top panel, $\delta^2$ and $\mathcal{F}_s^2 > 0$ cross over from negative to positive values almost simultaneously, indicating a transition from a regime where ZDMs and DMs coexist, to one with only ZDMs. In the bottom panel, both curves remain negative valued, indicating that both ZDMs and DMs are always present in the spectrum. Hence, $\delta^2$ and $\mathcal{F}_s^2$ effectively convey the same information about the boundary between the regime with only ZDMs and the regime where ZDMs and DMs coexist.}
    \label{fig:dsq_vs_fsq_kn}
\end{figure}

In the Kerr limit, Ref.~\cite{Yang:2013uba} showed numerically that for $2 \leq \ell \leq 100$, whenever $\mathcal{F}_{s}^2 > 0$, one also finds that $\delta^2 > 0$.
This property carries over to the \KN case in most situations.
To be precise about this statement, in Fig.~\ref{fig:dsq_vs_fsq_kn} we show the variation of $\delta^2$ and $\mathcal{F}_s^2$ over the extremal quarter circle,
$a_{\rm ext}^2 = \tfrac{1}{4} - q_{\rm ext}^2$,
for $s=-2$ and $(\ell, m) = (2,2)$ and $(2,1)$ in the top and bottom panels, respectively.
In the former case, as we increase the value of $a_{\rm ext}$, the functions $\delta^2$ and $\mathcal{F}_{-2}^2$
cross over to positive values at \emph{slightly} different values of the spin. In the latter case, $\delta^2$ and $\mathcal{F}_{-2}^2$
remain negative for all values of $a_{\rm ext}$.
We observed similar behaviors for different values of $s$, and multipoles $(\ell,m)$.
In conclusion, the two quantities $\delta^2$ and $\mathcal{F}_s^2$ effectively
convey the same information regarding the regime transition in the near-extremal \KN quasinormal mode spectrum.
Hence, we can use either quantity to determine where in the spin-charge
parameter space DMs begin to appear alongside the ZDMs; we chose $\delta^2$
because it has the advantage of being valid away from extremality, unlike
$\mathcal{F}^{2}_{s}$, which is valid strictly at extremality.

How does the $\delta^2$-criteria compare against the
$\mu$-criteria~\eqref{eq:summary_mu_criteria} derived in the WKB approximation?
To answer this question, we find it convenient to first introduce coordinates
$(\theta, \epsilon)$ to cover the spin-charge parameter space. We define them
as follows:
\begin{equation}
    a = (\tfrac{1}{2} - \epsilon)\cos{\theta},
    \quad \textrm{and} \quad
    q = (\tfrac{1}{2} - \epsilon)\sin{\theta}.
\label{eq:ethcoord}
\end{equation}
These definitions are motivated by the black-hole region of the spin-charge parameter space
being subject to the inequality $a^2 + q^2 \leq 1/4$.
We use $\theta$ to introduce the terminology ``high-rotation'' and
``high-charge'' limits, that loosely correspond to $\theta < 45^\circ$ and $\theta >
45^\circ$, respectively. Similarly to $\sigma$, the near-extremal regime corresponds to $\epsilon \ll 1$,
regardless of the value of $\theta$.

\begin{figure}[t]
\includegraphics{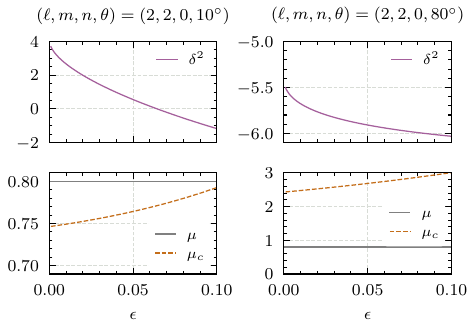}
\caption{Comparison between the $\delta^2$ and $\mu$-criteria for the fundamental $\ell = m = 2$ mode. In the left column we show
a high-spin case, $\theta = 10^{\circ}$, while in the right column we show a high-charge case, $\theta = 80^{\circ}$.
Close to extremality ($\epsilon \lesssim 10^{-3}$), when $\theta = 10^\circ$, $\delta^2 > 0$ and $\mu > \mu_c$, indicating a ZDM-only regime; when $\theta = 80^\circ$, $\delta^2 < 0$ and $\mu < \mu_c$, indicating a regime where ZDMs and DMs coexist.
Thus, for practical purposes, both criteria are equivalent.
}
\label{fig:dsqfixedtheta}
\end{figure}

\begin{figure}[t]
\includegraphics[width=\columnwidth]{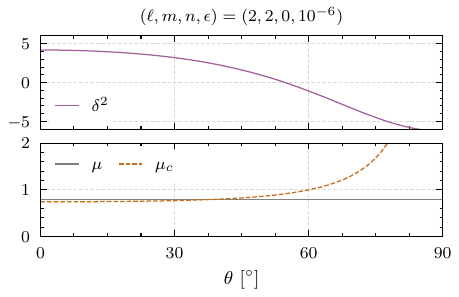}
\caption{Comparison between the $\delta^2$ and $\mu$-criteria for the fundamental $\ell = m = 2$ mode. The top panel shows $\delta^2$, while the bottom panel shows $\mu$ and $\mu_c$ as functions of $\theta$ for $\epsilon = 10^{-6}$.
As $\theta$ increases, both criteria predict a transition from a ZDM-only
regime to one where ZDMs and DMs coexist.
Moreover, both criteria predict similar values for the critical angle
$\theta_c$ at which the transition occurs: $\theta_c \approx 55^\circ$ using
the $\delta^2$ criteria and $\theta_c \approx 40^\circ$ using the $\mu$
criteria.
We see again that for practical purposes both criteria are equivalent.
}
\label{fig:dsqfixedevarytheta}
\end{figure}

We now compare the $\delta^2$ and
$\mu$-criteria by first fixing the angle $\theta$ to be $10^{\circ}$ (high-spin) and $80^{\circ}$ (high-charge), and varying $\epsilon$ from near extremality ($\epsilon = 10^{-6}$)
up to $\epsilon = 10^{-1}$. We sample the values of $\epsilon$ logarithmically,
and focus on the fundamental $\ell = m = 2$ mode.
We show the results of this comparison in
Fig.~\ref{fig:dsqfixedtheta}.
Close to extremality ($\epsilon \lesssim 10^{-3}$), we see that for $\theta = 10^\circ$ (left column), $\delta^2 > 0$ and $\mu > \mu_c$, indicating a ZDM-only regime. For $\theta = 80^\circ$ (right column), the same near-extremal region instead has $\delta^2 < 0$ and $\mu < \mu_c$, indicating the coexistence of ZDMs and DMs. Thus, for practical purposes, both the $\mu$ and $\delta^2$ criteria furnish equivalent predictions.
In Fig.~\ref{fig:dsqfixedevarytheta}, we repeat a similar exercise, but now keeping $\epsilon$ fixed at $10^{-6}$, and varying $\theta$ continuously between the high-spin and the high-charge limits.
We see that, as $\theta$ is increased, both criteria predict a transition from a ZDM-only regime to a regime where ZDMs and DMs coexist. Importantly, both predict nearly the same critical angle $\theta_c$ for the transition, indicating once again that, for practical purposes, the two criteria are equivalent.
Thus, the key outcome of this discussion is that we can generally use either the $\delta^2$ or the $\mu$-criteria to predict when ZDMs and DMs coexist. In the next section, we will test these predictions against
full numerical calculations of the \DF equation.

\begin{figure*}[ht]
    \includegraphics{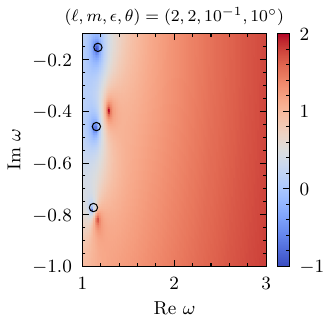}
    \includegraphics{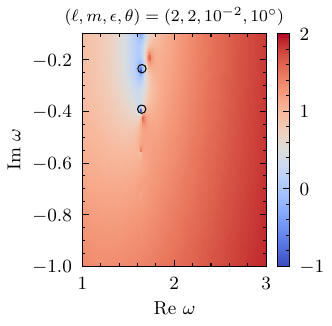}
    \includegraphics{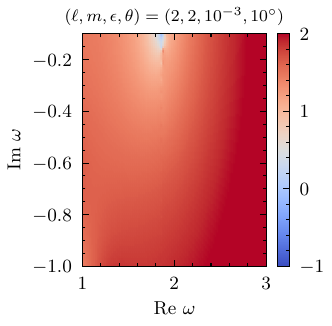}
\caption{Numerical evaluation of the logarithm of the radial continued fraction $|C_r|$, Eq.~\eqref{eq:cf_rad}, for the fundamental $(\ell,m)=(2,2)$ mode, for $\theta = 10^\circ$, and different values of $\epsilon$. Close to extremality ($\epsilon \lesssim 10^{-3}$), only the ZDMs exist, while further away from extremality the ZDMs transition into a set of modes with $|\mathrm{Im}\,\omega| > 0$, i.e., the DMs. This is consistent with the $\delta^2$ and $\mu$-criteria, since both $\delta^2$ and $\mu - \mu_c$ change signs from positive to negative as we move away from extremality (see Fig.~\ref{fig:dsqfixedtheta}).}
  \label{fig:gridsearch_fixedth10degvarye}
\end{figure*}

\begin{figure}[h!]
\includegraphics{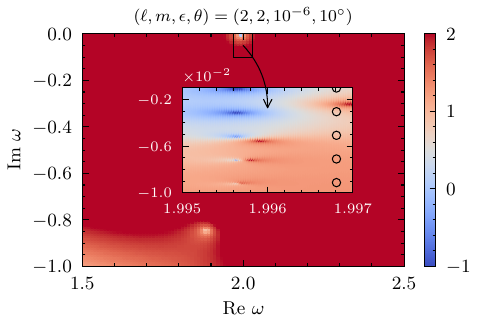}
\includegraphics{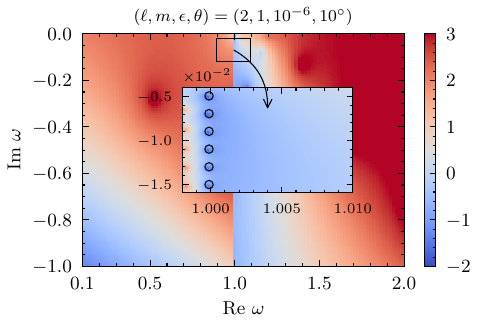}
\caption{Numerical evaluation of the logarithm of the radial continued fraction $|C_r|$, Eq.~\eqref{eq:cf_rad}, for the fundamental $(\ell,m)=(2,2)$ and $(2,1)$ modes, for $\theta = 10^\circ$ and $\epsilon = 10^{-6}$. We zoom in on the region of the grid containing the ZDMs, outlined by the rectangular boxes. The circles mark the predictions of Eq.~\eqref{eqref:zdmeqn} for the ZDMs, while the dark blue regions indicate where they would appear numerically.}
\label{fig:th10deg_zdmzoomed}
\end{figure}

\subsection{Zero-damped and damped modes near extremality: A numerical survey}

\begin{figure}[ht]
    \includegraphics[width=\columnwidth]{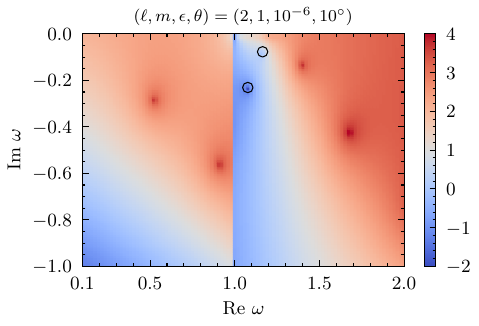}
    \caption{Numerical evaluation of the logarithm of the radial continued fraction $|C_r|$, Eq.~\eqref{eq:cf_rad}, for the fundamental $(\ell,m)=(2,1)$ mode with $\theta = 10^\circ$ and $\epsilon = 10^{-6}$. While Fig.~\ref{fig:th10deg_zdmzoomed} (bottom panel) illustrated the ZDMs for this case, we now shift our focus to the DMs, marked by circles. The ZDMs are still visible as the sharp feature at $\mathrm{Re}\,\omega \approx 1$. Taken together, these two figures demonstrate the coexistence of the ZDMs and DMs in this case, consistent with the predictions of both the $\delta^2$ and $\mu$ criteria, as $\delta^2 < 0$ and $\mu < \mu_c$.}
    \label{fig:lneqmgridsearch}
\end{figure}

Guided by the analytical results presented in Secs.~\ref{sec:wkb_criteria}
and~\ref{sec:dsqfsq}, we now investigate the quasinormal mode spectrum of the
near-extremal \KN spacetime, surveying numerically the existence of the ZDMs and DMs. To do so, we vary either $\epsilon$ or $\theta$ while keeping the other fixed, and perform a grid-search in the complex plane to locate the modes.

\begin{figure*}[hbt]
    \includegraphics{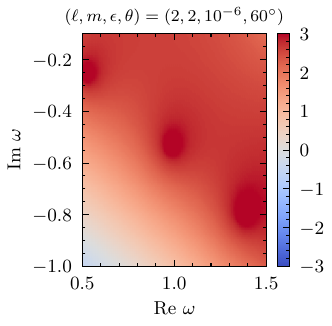}
    \includegraphics{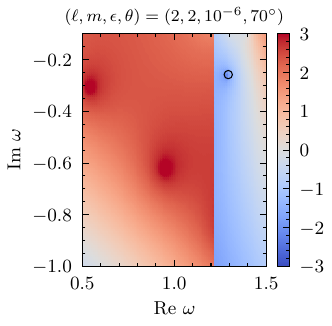}
    \includegraphics{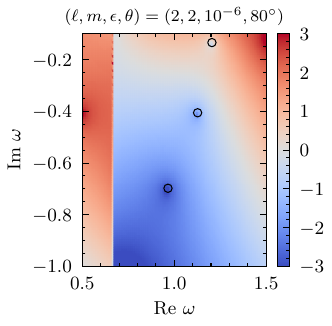}
    \caption{Numerical evaluation of the logarithm of Leaver's radial continued fraction $|C_r|$, for the fundamental $(\ell,m)=(2,2)$ mode, for $\epsilon = 10^{-6}$ and different values of $\theta$. There are no DMs for $\theta = 60^\circ$; they only appear for $\theta \gtrsim 70^\circ$ (marked by circles). This behavior is consistent with the $\delta^2$- and $\mu$-criteria, since both $\delta^2$ and $\mu - \mu_c$ change sign from positive to negative as $\theta$ is increased, i.e., as we move from the high-spin to the high-charge limit (see Fig.~\ref{fig:dsqfixedevarytheta}).}
\label{fig:gridsearch_fixede1m6varyth}
\end{figure*}

We first fix $\theta$ to be $10^\circ$(high-rotation limit), and increase $\epsilon$ from $10^{-6}$ to $10^{-1}$ logarithmically.
For each value of $\epsilon$, we calculate $\log_{10}\left|C_r\right|$ on a $100
\times 100$ grid in the complex-$\omega$ plane, where $C_r$ is the radial
continued fraction~\eqref{eq:cf_rad}.
Our findings are summarized as follows:
\begingroup
\allowdisplaybreaks
\begin{enumerate}
    \item In Fig.~\ref{fig:gridsearch_fixedth10degvarye}, we show our results for the $(\ell, m) = (2,2)$ mode with $\theta = 10^\circ$.
    Far from extremality ($\epsilon \gtrsim 10^{-3}$), we identify only a single branch of quasinormal modes with $\rm|Im\,\omega| > 0$ (left and middle panels).
    Closer to extremality ($\epsilon \lesssim 10^{-3}$), these
    modes transition into ZDMs, with $\rm|Im\,\omega| \rightarrow 0$ (right panel).
    For instance, when $\epsilon = 10^{-3}$, the ZDMs are localized at $\textrm{Re}\,\omega \approx 1.92$.
    Since the individual ZDMs cannot be resolved by the grid resolution in Fig.~\ref{fig:gridsearch_fixedth10degvarye},
    we confirm their existence in Fig.~\ref{fig:th10deg_zdmzoomed} (top panel) for the specific case of $\epsilon = 10^{-6}$. In particular, we perform a grid search over the smaller region of the complex plane where the ZDMs are localized.
    The main takeaway is that for $\theta = 10^\circ$ and $(\ell,m) = (2,2)$, the near-extremal spectrum does not contain any DMs,
    consistent with the predictions of both the $\delta^2$ and $\mu$-criteria,
    since both $\delta^2>0$ and $\mu > \mu_c$; see
    Fig.~\ref{fig:dsqfixedtheta}.
    \item We then repeat the same analysis for $(\ell,m) = (2,1)$, with $\theta
    = 10^\circ$. In this case, both ZDMs and DMs exist in the near-extremal limit.
    For $\epsilon = 10^{-6}$, the ZDMs are shown in Fig.~\ref{fig:th10deg_zdmzoomed} (bottom panel), while the DMs are shown in Fig.~\ref{fig:lneqmgridsearch}. Once again, this behavior is consistent with the predictions of both the $\delta^2$ and $\mu$-criteria, since in this case $\delta^2<0$ and $\mu < \mu_c$.
\end{enumerate}
\endgroup

We now fix $\epsilon = 10^{-6}$, thereby restricting ourselves to the near-extremal region, and perform similar grid searches as we vary $\theta$ from $60^\circ$ to $80^\circ$, that is, in the neighborhood
of the predicted critical angle $\theta_{c}$ where we expect the transition to occur. See the discussion around Fig.~\ref{fig:dsqfixedevarytheta}.
We show the outcome of these grid searches in
Fig.~\ref{fig:gridsearch_fixede1m6varyth}. We see that the DMs only appear for
large values of $\theta$, i.e., in the high-charge limit. Yet again, this is
consistent with the predictions of both the $\delta^2$ and $\mu$-criteria,
since both $\delta^2$ and $\mu - \mu_c$ change sign from positive to negative
as $\theta$ increases for $\epsilon = 10^{-6}$; see
Fig.~\ref{fig:dsqfixedevarytheta}.

\section{Near-horizon--photon-sphere modes and their connection to (zero-)damped modes} \label{sec:nh_ps_zdm_dm_connection}

The works by Dias~et~al.~\cite{Dias:2015wqa,Dias:2021yju,Carullo:2021oxn,Dias:2022oqm}
found that the \KN quasinormal mode spectrum exhibits two families of modes, the PS and the NH modes. Physically, the PS modes were shown to be closely related to the properties of unstable, equatorial photon orbits in the \textit{eikonal limit}, although this interpretation was found to be reasonably valid even for modes with small $\ell$ \cite{Dias:2022oqm}. Meanwhile, the NH modes were shown to have wave functions that were highly localized near the horizon \cite{Dias:2022oqm}. An additional complication that arose was that the distinction between the two types of modes was sharp only when $a/q \ll 1$, i.e., in the \RN limit. Beyond this limit, the two families were found to blend together into a single family of modes, reminiscent of the ``eigenvalue repulsion'' phenomenon in quantum mechanics~\cite{Dias:2022oqm}. Hence, they are best referred to as a composite NH-PS family of modes.

In this section, we first review some analytical results
from Ref.~\cite{Dias:2022oqm}; see Eqs.~\eqref{eq:nhpsmaeformula}
and~\eqref{eq:wps_wkb_finalexpr}.
Using these results, we then attempt to draw some connections between the composite NH-PS
modes and the ZDMs and DMs that we have encountered so far.
We continue to omit the subscripts $\ell$,
$m$, $n$, and $s$ in our equations to keep the notation light.

\subsection{Expression for the dominant near-horizon--photon-sphere mode near extremality} \label{sec:nhps_formula}
We now reproduce an analytic formula from Ref.~\cite{Dias:2022oqm}, which gives the frequency of the dominant mode for the full \KN solution in the near-extremal limit. Shown below, this expression is derived via a matched asymptotic expansion similar to that used to obtain Eq.~\eqref{eqref:zdmeqn} for the ZDM frequencies:
\begin{equation}
    \begin{split}
        \omega_{\rm NH-PS} &= \frac{2 ma_{\text{ext}}}{1 + a_{\text{ext}}^2} + \sigma \left[\frac{ma_{\text{ext}}(1 - 4a_{\text{ext}}^2)}{(1 + 4a_{\text{ext}}^2)^2} \right.
        \\
        &\left. \quad - \frac{\rm {i}}{4} \frac{1 + 2n}{1 + 4a_{\text{ext}}^2} - \frac{\sqrt{-\lambda_2(m,2a_{\text{ext}})}}{4(1+4a_{\text{ext}}^2)^2} \, \right],
    \end{split}
\label{eq:nhpsmaeformula}
\end{equation}
where $a_{\text{ext}}^2 = 1/4 - q_{\text{ext}}^2$, and $\lambda_2$ is given by a WKB expansion:
\begin{equation}
    \lambda_2 = \lambda_{2,0}m^2 + \lambda_{2,1}m + \lambda_{2,2} + \frac{\lambda_{2,3}}{m} + \mathcal{O}(1/m^2).
\end{equation}
The coefficients $\lambda_{2,i}$, $i \in \mathbb{N}$, can be found in Ref.~\cite{Dias:2022oqm}, Eqs.~(3.35a-c).
Equation~\eqref{eq:nhpsmaeformula} was originally derived for the NH modes that dominate the near-extremal spectrum in the high-charge limit. However, it turns out that it also describes the dominant composite NH-PS mode frequencies near extremality, \textit{irrespective} of the value of $a/q$. Moreover, although the derivation assumed the eikonal limit $\ell \gg 1$, Eq.~\eqref{eq:nhpsmaeformula} was found to be an excellent approximation to numerical results even for small values of $\ell$.

\subsection{Equatorial photon-sphere modes: WKB analysis} \label{sec:ps_eikonal}

Next, we introduce another useful expression, also derived in Ref.~\cite{Dias:2022oqm}, for the frequencies of the $\ell = |m|$ equatorial PS modes in the eikonal limit:
\begin{align}
    \omega_{\rm PS} \approx \frac{m}{b_\sr} - \ii \frac{n + 1/2}{b_\sr r_\sr^2} \frac{|r_\sr^2 + a^2 - ab_\sr|}{|b_\sr - a|} \sqrt{6r_\sr^2 + a^2 - b_\sr^2}\,,
    \label{eq:wps_wkb_finalexpr}
    \nonumber \\
\end{align}
where $r_\sr$ and $b_\sr$ denote the radius and impact parameter of the photon
orbit. These quantities are determined by the following system of algebraic equations,
\begin{equation}
    \begin{split}
         - \left\{
             \tfrac{3}{2}r_\sr
             +\left[\tfrac{9}{4}r_\sr^2-8q^2(r_\sr^2+2a(\Delta^{1/2}_\sr + a )) \right]^{1/2}
             \right\}^2\\
         + 4 \left[ r_\sr^2 + 2a(\Delta^{1/2}_\sr + a) \right]^2 = 0,
    \end{split}
    \label{eq:wps_wkb_rs}
\end{equation}
and
\begin{equation}
    b_\sr = \frac{r_\sr^2 \, \Delta^{1/2}_\sr + a(q^2 - r_\sr)}{r_\sr^2 - r_\sr + q^2},
    \quad \textrm{where} \quad
    \Delta_\sr = \Delta(r_\sr).
    \label{eq:wps_wkb_bs}
\end{equation}
Equation~\eqref{eq:wps_wkb_bs} yields two independent solutions, denoted by $r_\sr^{+}$ and $r_\sr^{-}$, which correspond to the orbital radii for the corotating ($m>0$) and counter-rotating ($m<0$) modes, respectively.
The expression~\eqref{eq:wps_wkb_finalexpr} turns out to be an excellent approximation to numerical results even for small values of $\ell = \left|m\right|$, although it is strictly valid only in the eikonal limit~\cite{Dias:2022oqm}.

\subsection{Relating the near-horizon--photon-sphere modes and the zero-damped and damped modes}
\label{sec:relation_nhps_zdms}

We now establish some connections between the composite NH-PS modes of the full \KN solution, and the ZDMs and DMs within the \DF approximation.
Our discussion generally focuses on the $(\ell, m, n) = (2, 2, 0)$ mode, but the
conclusions that we draw for this mode are also shared with the $(2,2,1)$ and
$(3,3,0)$ modes.

\begin{figure}[htb]
\includegraphics{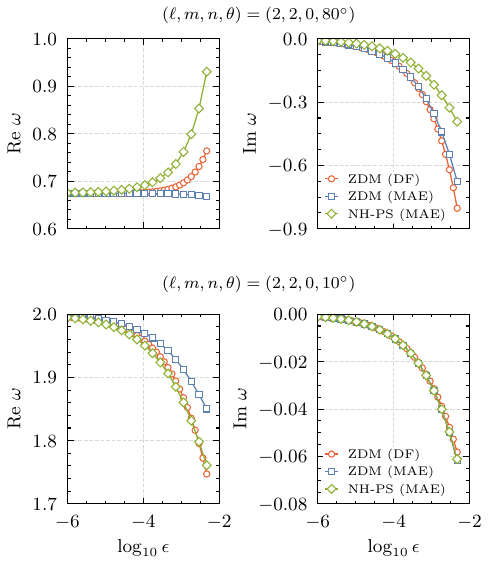}
\caption{Predictions of the matched-asymptotic expansion (``MAE'') for the ZDMs~\eqref{eqref:zdmeqn} and for the composite NH-PS modes~\eqref{eq:nhpsmaeformula} as functions of $\epsilon$ for fixed $\theta = 80^\circ$ (top panel) and $\theta = 10^\circ$ (bottom panel). We also show a numerical branch of ZDMs seeded with either expression at $\epsilon = 10^{-6}$, and tracked away from extremality by solving the \DF (``DF'') equation thereafter. All three curves converge to ${\rm Re}\,\omega \approx m\Omega_H^{\text{ext}}$ and ${\rm Im}\,\omega \approx 0$ as $\epsilon \rightarrow 0$, i.e., as extremality is approached.}
\label{fig:zdm_vs_nhps_varyeps_th}
\end{figure}

We begin with a direct comparison between the analytic expressions for the
ZDM and composite NH-PS mode frequencies, given by Eqs.~\eqref{eqref:zdmeqn}
and~\eqref{eq:nhpsmaeformula}, respectively.
In Fig.~\ref{fig:zdm_vs_nhps_varyeps_th} we track the predictions of these expressions as $\epsilon$ is varied for fixed values of $\theta$.
In the same figure, we also show a numerical
branch seeded with either expression at $\epsilon = 10^{-6}$, and tracked
away from extremality thereafter.
Close to extremality, Eqs.~\eqref{eqref:zdmeqn}
and~\eqref{eq:nhpsmaeformula} agree with the numerical branch, as well as with each other, which is unsurprising since they are identical to leading order in the off-extremal parameter $\sigma$.
Farther from extremality, the two expressions generally disagree with each other, as well as with the numerical branch.

\begin{figure}
\includegraphics{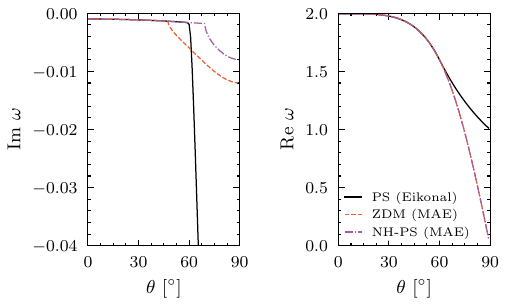}
\caption{Predictions of Eqs.~\eqref{eqref:zdmeqn}, \eqref{eq:nhpsmaeformula}, and \eqref{eq:wps_wkb_finalexpr} (dashed, dot-dashed, and solid curves), for the ZDM and composite NH-PS mode frequencies obtained using the matched-asymptotic expansion technique, and the $\ell = |m|$ eikonal PS mode frequencies, respectively, as functions of $\theta$ at fixed $\epsilon = 10^{-6}$.
For the real part of $\omega$, the dashed and dot-dashed curves overlap.
In this near-extremal limit, Eqs.~\eqref{eqref:zdmeqn} and \eqref{eq:nhpsmaeformula} always coincide with each other, while they only coincide with Eq.~\eqref{eq:wps_wkb_finalexpr} for small values of $\theta$ i.e, when the spin $a$ of the black hole is larger than its charge $q$. Thus, in the near-extremal, high-charge limit, the ZDMs correspond to the NH modes, whereas in the near-extremal, high-rotation limit, they correspond to the PS modes, i.e., the composite NH-PS family of modes.}
\label{fig:ps_vs_nhpszdm}
\end{figure}

Next, we restrict ourselves to the near-extremal region by fixing $\epsilon = 10^{-6}$, and then compare Eq.~\eqref{eqref:zdmeqn} [or equivalently, Eq.~\eqref{eq:nhpsmaeformula}] to Eq.~\eqref{eq:wps_wkb_finalexpr} while varying $\theta$. That is, we compare the ZDMs with the eikonal PS modes near extremality, moving between the high-rotation and the high-charge limits. The results of this comparison are shown in Fig.~\ref{fig:ps_vs_nhpszdm}, indicating that Eq.~\eqref{eq:wps_wkb_finalexpr} agrees with Eqs.~\eqref{eqref:zdmeqn} and \eqref{eq:nhpsmaeformula} \textit{only} for small $\theta$, i.e. in the high-rotation limit. This implies that near extremality, ZDMs correspond to PS modes in the \textit{high-rotation} limit, and to NH modes in the \textit{high-charge} limit.

\begin{figure}
\includegraphics[width=\columnwidth]{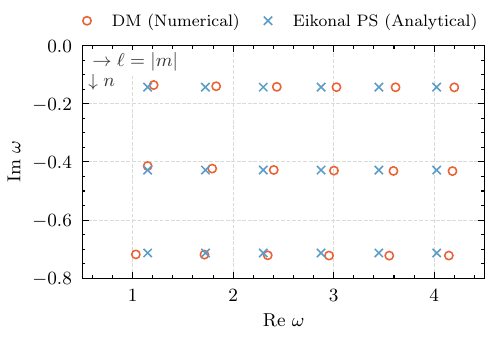}
\caption{The fundamental mode and the first two overtones of the eikonal PS modes given by Eq.~\eqref{eq:wps_wkb_finalexpr} (crosses), and the numerically calculated gravitational DMs, using Eq.~\eqref{eq:wps_wkb_finalexpr} as the initial seed (circles), for $(\epsilon,\theta) = (10^{-6},80^\circ)$ and $\ell = \left|m\right|=\{2,\ldots,7\}$. There is a clear correspondence between the two sets of modes, which means that in the near-extremal, high-charge limit, the $\ell = |m|$ DMs are equivalent to the PS modes. We also verified separately that Eq.~\eqref{eq:wps_wkb_finalexpr} is accurate up to $\mathcal{O}(1)$ and $\mathcal{O}(1/\left|m\right|)$ for the real and imaginary parts of the frequencies, respectively.}
\label{fig:eikps_vs_dm_highcharge}
\end{figure}

This raises an interesting follow-up question: since the NH and PS modes are \textit{distinct} in the near-extremal, high-charge limit (unlike in the high-rotation limit), if the ZDMs in this regime correspond to the NH modes, do the DMs then correspond to the PS modes? We explore this further in Fig.~\ref{fig:eikps_vs_dm_highcharge}, where we show the fundamental and first two overtones of the eikonal PS mode frequencies given by Eq.~\eqref{eq:wps_wkb_finalexpr}, and the numerically calculated gravitational DMs (with Eq.~\eqref{eq:wps_wkb_finalexpr} as the initial seed) for $(\epsilon,\theta) = (10^{-6},80^\circ)$ and $\ell = \left|m\right|=\{2,\dots,7\}$. From this figure, we see clear agreement between the two sets of modes, providing strong evidence that the DMs align with the PS modes in this regime. We also verified that our numerical results are consistent with Eq.~\eqref{eq:wps_wkb_finalexpr} being accurate to $\mathcal{O}(1)$ and $\mathcal{O}(1/\left|m\right|)$ for the real and imaginary parts of the frequencies, respectively \cite{Dias:2022oqm}.

With this, we conclude our discussion regarding the connections between the ZDMs/DMs and the NH-PS modes. While the insights presented here are valuable, one should exercise caution when extending them to more general cases, such as modes with $\ell \neq m$, since the eikonal PS expression~\eqref{eq:wps_wkb_finalexpr} is valid only for the $\ell = |m|$ modes.

\section{Highly damped modes} \label{sec:highly_damped_df_qnms}

\begin{figure*}
\includegraphics[width=\columnwidth]{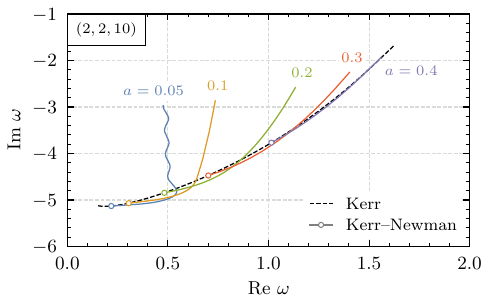}
\includegraphics[width=1.02\columnwidth]{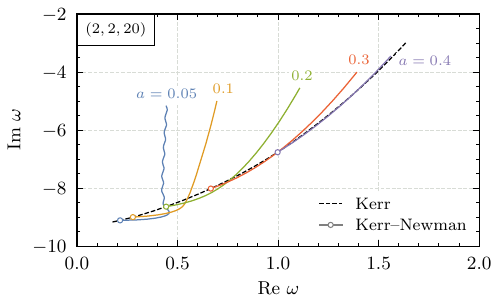}
\\
\includegraphics[width=\columnwidth]{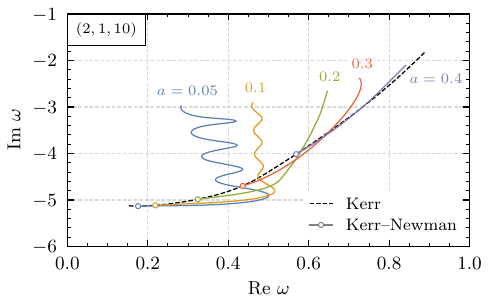}
\includegraphics[width=1.02\columnwidth]{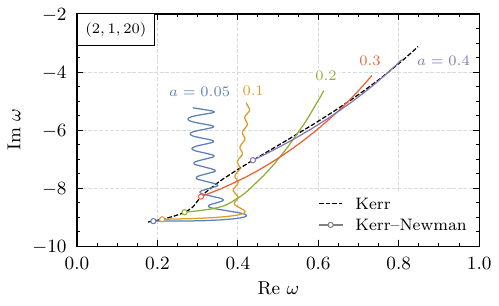}
\caption{Trajectories of the $\ell = 2$, $m>0$ Kerr and \KN gravitational QNMs, as extremality is approached. The dashed black lines show the evolution of the Kerr modes as the rotation $a$ of the black hole is increased. The solid lines (beginning at the circular markers) show the evolution of the \KN modes as the charge $q$ of the black hole is increased, with the rotation $a$ held fixed at several different values (indicated by the color of the curves).
Our calculations are done up to $\epsilon \approx 0.05$.
The overall ``shapes'' of the $q \neq 0$ mode trajectories are largely determined by the azimuthal index $m$, with changes to the overtone index $n$ introducing only minor variations.
Additionally, the branches with smaller values of $a$ exhibit a pronounced oscillatory structure, while those with higher values of $a$ asymptote to the Kerr branch.}
\label{fig:branchplots_mpositive}
\end{figure*}

We now investigate the highly-damped gravitational quasinormal modes of \KN
black holes in the \DF approximation, with the goal of better understanding the
asymptotic, large overtone number $n$ regime of the spectrum.
Following Ref.~\cite{Berti:2003jh}, which conducted a similar study for the
Kerr solution, we define the (approximate) onset of the asymptotic regime
relative to the purely imaginary ``algebraically special mode'' in the
Schwarzschild quasinormal mode spectrum.
We do so because this mode effectively separates the
spectrum into upper and lower branches, with the latter moving rapidly downward
in the complex plane with increasing overtone number $n$.
The overtone number $n$ corresponding to this mode
increases rapidly with $\ell$; for $\ell = 2$, it occurs at $n = 8$, whereas for
$\ell = 3$, it occurs at $n = 41$ (see Leaver~\cite{Leaver:1985ax}, Fig.~1).
In practice, it becomes increasingly challenging for our root-finder to converge for
overtone numbers beyond $n \gtrsim 40$, a difficulty also encountered in
Ref.~\cite{Berti:2003jh}.
For this reason, we focus only on the $\ell = 2$ modes.
We remark that the large-$n$ quasinormal mode spectrum was also studied in the
Schwarzschild limit in Refs.~\cite{Nollert:1993zz,Andersson:1992scr,Motl:2002hd,Motl:2003cd,Andersson:2003fh,Natario:2004jd}, in the \RN limit in Refs.~\cite{Andersson:1996xw,Berti:2003zu}, and
in the Kerr limit in Refs.~\cite{Keshet:2007nv,Kao:2008sv,Daghigh:2010wm}.

\begin{figure*}[ht]
\includegraphics{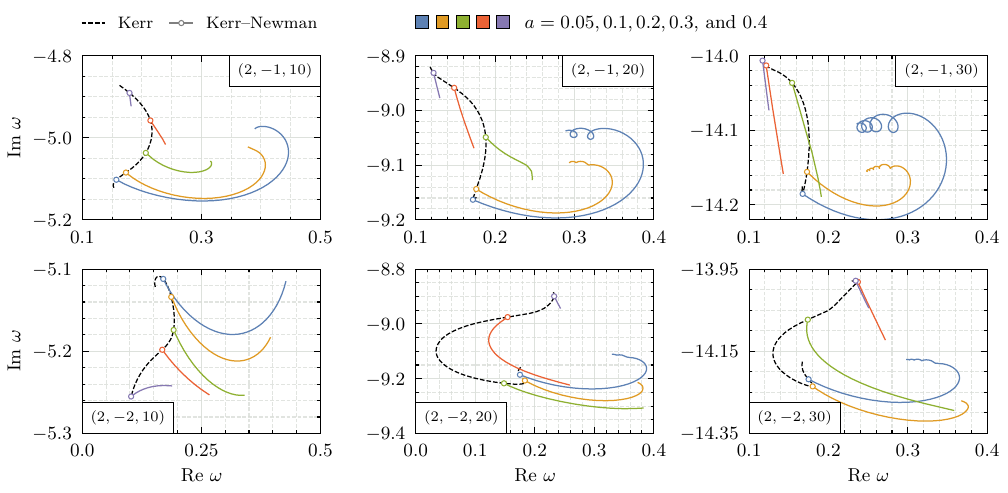}
\caption{Trajectories of the $\ell = 2$, $m<0$ Kerr and \KN gravitational QNMs, as extremality is approached. The dashed black lines show the evolution of the Kerr modes as the rotation $a$ of the black hole is increased. The solid lines (beginning at the circular markers) show the evolution of the \KN modes as the charge $q$ of the black hole is increased, with the rotation $a$ held fixed at several different values (indicated by the color of the curves).
Our calculations are done up to $\epsilon \approx 0.05$.
In this case, the shapes of the $q \neq 0$ mode trajectories are less amenable to classification with respect to the azimuthal index $m$, unlike the $m>0$ modes shown in Fig.~\ref{fig:branchplots_mpositive}.
In particular, for a given value of $m$, the overtone index $n$ now plays a significant role in defining the shape of the curve.
For instance, with regards to the $m = -1$ modes, branches with smaller values of $a$ exhibit a ``spiraling'' behavior that becomes remarkably more pronounced as $n$ is increased. On the other hand, while the $m=-2$ modes also exhibit a similar behavior, variations with the overtone index $n$ are much less pronounced than for $m=-1$.}
\label{fig:branchplots_mnegative}
\end{figure*}

\begin{figure}[]
\includegraphics{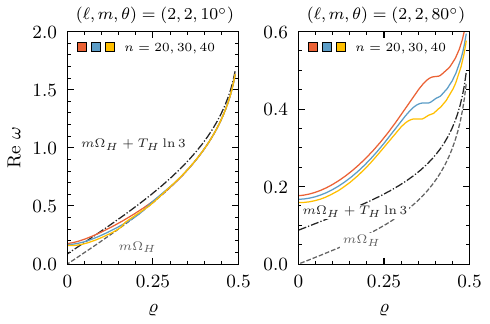}
\caption{The real part of the quasinormal frequencies of the $(\ell,m) = (2,2)$ gravitational modes, for $\theta = 10^\circ$ (left panel) and $\theta = 80^\circ$ (right panel) and $\varrho = \sqrt{a^2 + q^2} \in [0,0.5)$. We show the numerically computed modes for $n = 20$, $30$, and $40$, together with Hod's conjecture~\eqref{eq:hodsconj} (dot-dashed curves), and the simplified formula~\eqref{eq:hodsconj_simpler} (dashed curves). Note that we do not track the curves all the way up to extremality, $\varrho = 0.5$, as it becomes increasingly difficult to reliably find the quasinormal frequencies in this limit.
For $\theta = 10^\circ$, the curves agree with each other quite well all the
way up to the near-extremal limit, while for $\theta = 80^\circ$, the agreement
is generally poor.}
\label{fig:fixth_varyeps_m2}
\end{figure}

\begin{figure}[]
\includegraphics{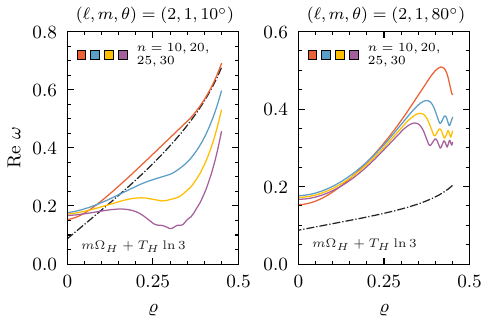}
\caption{The real part of the quasinormal frequencies of the $(\ell,m) = (2,1)$ gravitational modes, for $\theta = 10^\circ$ (left panel) and $\theta = 80^\circ$ (right panel) and $\varrho = \sqrt{a^2 + q^2} \in [0,0.5)$. We show the numerically computed modes for $n = 10$, $20$, $25$, and $30$, together with the predictions of Hod's conjecture~\eqref{eq:hodsconj}.
As in Fig.~\ref{fig:fixth_varyeps_m2}, and for the same reason, we do not track the curves all the way up to extremality, $\varrho = 0.5$.
The predictions of Eq.~\eqref{eq:hodsconj} do not agree with the numerical
results at all, unlike in Fig.~\ref{fig:fixth_varyeps_m2}, where we saw
agreement for the case where $\theta = 10^\circ$.}
\label{fig:fixth_varyeps_m1}
\end{figure}

We begin our study of the large-$n$ limit of the \KN quasinormal mode spectrum as follows.
First, we compute the quasinormal frequencies in the Kerr limit, increasing the spin from zero
up to near extremality, for fixed values of $\ell$, $m$, and $n$.
Then, for a few values of the spin $a$ along this curve, we increase the charge $q$ up to the near-extremal limit, where
$a^2 + q^2 = (1/2 - \epsilon)^2$; cf.~Eq.~\eqref{eq:ethcoord}.
Our ability to approach extremality is constrained by the numerical challenges that we mentioned earlier; in our calculations, we were able to reach $\epsilon \approx 0.05$ and $\epsilon \approx 0.1$ for the $m>0$ and $m<0$ modes, respectively, using $N = 10^3$ terms in the radial continued fraction.
We validated our choice of $N$ by comparing the Kerr $(\ell, m, n) = (2,1,30)$ mode frequencies within the range $a \in [0,0.49]$, which we covered with 5001 evenly spaced values. We computed the frequencies using $N$ and $5N$ terms in the radial continued fraction, and took the absolute difference between the two calculations to quantify our errors. By doing so, we estimated our errors to be of the order of $\approx 10^{-4}$ for both the real and imaginary parts of the frequencies.

The ``Kerr trajectory,'' along with various $q \neq 0$ (\KN) ``branches,'' are shown in Figs.~\ref{fig:branchplots_mpositive} and~\ref{fig:branchplots_mnegative}, for numerous combinations of $(\ell,m,n)$-values. Based on these plots, we observe the following trends:
\begin{enumerate}
    \item The branches that emerge at small values of $a$ exhibit a distinctive behavior: oscillatory along the real axis for $m>0$, and spiraling trajectories for $n \geq 20$ when $m<0$. See the trajectories labeled $a = 0.05$ and $a=0.1$ in Figs.~\ref{fig:branchplots_mpositive} and~\ref{fig:branchplots_mnegative}. These behaviors disappear when the branching occurs at larger values of $a$.
    \item For the $m>0$ modes, the overall shape of the trajectory appears to be largely determined by the azimuthal index $m$, with the overtone number $n$ introducing only minor variations; compare the panels across the two rows in Fig.~\ref{fig:branchplots_mpositive}.
    \item The $m<0$ modes seem to be more sensitive to the overtone number $n$; for instance, the spiraling behavior of the $m=-1$ modes is significantly affected by the specific choice of $n$.
Note how the number of whirls increases as we go from overtone number $n=20$ to $n=30$ in the top row of Fig.~\ref{fig:branchplots_mnegative}.
\end{enumerate}

We now turn to an investigation of the validity of two expressions shown below
\begin{equation}
        \textrm{Re}\, \omega = T_H \ln{3} + m\Omega_H,
        \label{eq:hodsconj}
\end{equation}
and
\begin{equation}
        \textrm{Re}\,\omega = m\Omega_H.
        \label{eq:hodsconj_simpler}
\end{equation}
In both equations, $\Omega_H$ is the angular frequency of the horizon [cf.~Eq.~\eqref{eq:omega_H_extremal}],
and
\begin{equation}
    T_H = \frac{1}{4\pi}\frac{r_+ - r_-}{r_+ - q^2},
\end{equation}
is the temperature of the black hole. We remark that, as observed in Ref.~\cite{Berti:2004um}, Eq.~\eqref{eq:hodsconj_simpler} is a good representation of the $m>0$ modes only in an
``intermediate asymptotic'' regime. As $n \to \infty$, the real part of $\omega$ is
no longer given by a simple polynomial function involving $T_H$ and $\Omega_H$.

Our motivations for doing this comparison are twofold.
First, Eq.~\eqref{eq:hodsconj}, sometimes known as ``Hod's conjecture,'' was originally proposed for the asymptotic frequencies of the Kerr quasinormal modes~\cite{Hod:1998vk}, motivated by an application of Bohr's correspondence principle to black-hole thermodynamics. See, for e.g., Refs.~\cite{Dreyer:2002vy,Maggiore:2007nq}.
However, it was later shown that this equation does \textit{not} hold for general $(\ell, m)$ modes~\cite{Berti:2003jh}. As such, we do not expect Eq.~\eqref{eq:hodsconj} to hold in the \KN case either. This means that the comparison between our numerical results and Eq.~\eqref{eq:hodsconj} serves to explicitly highlight this discrepancy.
Second, Ref.~\cite{Berti:2003jh} observed that even though Eq.~\eqref{eq:hodsconj} is incorrect, Eq.~\eqref{eq:hodsconj_simpler} provides a very good approximation for the real parts of the highly-damped $\ell = m = 2$ frequencies of Kerr black holes.
The lack of any similar large-$n$ calculations for \KN black holes motivates us
to examine the validity of Eq.~\eqref{eq:hodsconj_simpler} in this context,
even though we are limited to the \DF approximation.

In Figs.~\ref{fig:fixth_varyeps_m2} and \ref{fig:fixth_varyeps_m1}, we show $\textrm{Re}\,\omega$
for modes with $\ell = 2$, azimuthal indices $m=2$ and $m=1$, and various overtone numbers $n$ indicated in the legends.
Specifically, we track the evolution of  $\textrm{Re}\,\omega$ as we increase $\varrho =
(a^2 + q^2)^{1/2}$ from zero up to the near-extremal limit, for $\theta = 10^\circ$ and
$\theta = 80^\circ$.
The angular frequency of the horizon, expressed in terms of $\varrho$, is:
\begin{equation}
    \Omega_H = \frac{4a}{4a^2 + (1 + \sqrt{1 - 4\varrho^2})^2}.
\end{equation}
We use this relation to also plot Eqs.~\eqref{eq:hodsconj} and \eqref{eq:hodsconj_simpler}.
We show the latter equation in Fig.~\ref{fig:fixth_varyeps_m2} in which $\ell = m = 2$.

Unsurprisingly, we find that Eq.~\eqref{eq:hodsconj} remains a poor approximation
in the \KN case.
On the other hand, Eq.~\eqref{eq:hodsconj_simpler}, proposed in the context of
the Kerr solution, provides a good approximation to our numerical results for
$\theta = 10^\circ$ and $\varrho \gtrsim 0.2$. This corresponds to values $\Omega_H \gtrsim 0.2$.
However, for $\theta = 80^\circ$, Eq.~\eqref{eq:hodsconj_simpler} is no longer a good
representation of the numerical results.
In general, the numerical results for the $\ell = m$ modes exhibit \textit{nonmonotonic} dependence on $\varrho$, indicating that the true frequency behavior cannot be captured via a linear relationship with $\Omega_H$.

\section{Conclusions} \label{sec:conclusions}
We presented an extensive study of the quasinormal mode spectrum of the \KN black hole under the \DF approximation. First, we carried out a quantitative study of the validity of the approximation by comparing it to the full \KN quasinormal mode spectrum, as given in Refs.~\cite{Dias:2015wqa,Dias:2021yju,Carullo:2021oxn,Dias:2022oqm}. Using a combination of earlier numerical results and Bayesian fitting techniques, we assessed the accuracy of the \DF approximation over various regions of parameter space, including the $a = q$ line, and more generally the black hole parameter space defined by $a^2 + q^2 \leq 1/4$. We found the errors to be somewhat mode-dependent, generally staying approximately below $10 \%$ for the real parts and $1\%$ for imaginary parts. The largest errors were found to occur in the high-charge, near-extremal regime. This likely reflects the breakdown of the \DF approximation at large values of the black hole charge $q$, where neglecting the coupling between electromagnetic and gravitational perturbations is unjustified~\cite{Berti:2005eb}. A further source of error may be the omission of the NH modes in the fitting procedure~\cite{Carullo:2021oxn}, which dominate the spectrum in certain regions of the near-extremal parameter space.

Next, we investigated the \KN quasinormal mode spectrum in the near-extremal regime within the \DF approximation. We confirmed the existence of ZDMs for gravitational perturbations, extending earlier results that focused on scalar perturbations~\cite{Zimmerman:2015trm}. We also derived analytic conditions defining the boundaries between regions in spin-charge parameter space containing only ZDMs and those containing both ZDMs and DMs.
We also briefly discussed the two families of modes constituting the full \KN spectrum, known as the PS and NH modes, and highlighted their connections to the ZDMs and DMs of the \DF equation with $\ell = \left|m\right|$. Near extremality, we showed that in the \textit{high-charge} limit, the ZDMs and DMs correspond to the NH and PS modes, respectively, while in the \textit{high-rotation} limit (where no DMs exist), the ZDMs correspond to the composite NH-PS family of modes.

Finally, we investigated, for the first time, the asymptotic regime of the \KN
quasinormal mode spectrum, focusing on gravitational modes with $\ell = 2$.
We first studied the trajectories of the modes in the complex plane as the
charge $q$ was increased toward extremality, with the rotation $a$ held fixed
at several different values; see~Figs.~\ref{fig:branchplots_mpositive} and~\ref{fig:branchplots_mnegative}.
We found that for the $m>0$ modes the trajectories were largely dependent on the value
of $m$, while modes with $m<0$ had trajectories with an additional dependence on the
overtone number $n$.
We also studied the accuracy of two analytic expressions for the real part of
the quasinormal frequencies in the large-$n$ limit, namely ``Hod's
conjecture''~\eqref{eq:hodsconj} and its simplified
version~\eqref{eq:hodsconj_simpler}.
The former was proposed to hold for all $(\ell, m)$ modes and any black-hole
solution~\cite{Hod:1998vk}, whereas the latter was proposed for the $\ell = m =
2$ mode of Kerr~\cite{Berti:2003jh}.
We found that Eq.~\eqref{eq:hodsconj_simpler} provides an accurate description
of the real part of the modes (as described by the \DF equation) even when $q$
is nonzero but small, i.e., in the small $\theta$ limit.

We conclude by discussing some possibilities for future work. First, it would be interesting to investigate the relationship between the PS modes and DMs with $\ell \neq |m|$, extending our present analysis where we only study modes with $\ell = |m|$.
Second, on the numerical side, one could try to extend the large-$n$ computations to overtone numbers significantly higher than those considered in our current work. One way to do so would be to generalize the technique of Ref.~\cite{Berti:2004um} for Kerr black holes to the \DF equation for \KN black holes. This is important because, as noted in Ref.~\cite{Berti:2004um}, for the Kerr $\ell = m = 2$ mode, overtone numbers $n \gtrsim 40$ represent only an ``intermediate'' regime rather than the truly asymptotic limit; a similar situation can reasonably be expected for the corresponding \KN modes. One could also go further and obtain the large-$n$ overtones for the full \KN problem, and compare those results with the ones presented here in the \DF approximation.
Third, it could be worthwhile to study the properties of ZDMs in the context of extensions to general relativity, for two main reasons.
The first one is that it was recently shown that near-extremal black holes are particularly sensitive to higher-derivative corrections to general relativity~\cite{Horowitz:2023xyl,Horowitz:2024dch}; since ZDMs only emerge near extremality, they could serve as a valuable probe for studying such corrections.
The second reason is that because these modes are long lived, any imprint of physics beyond general relativity could, in principle, persist longer within these modes and therefore be potentially easier to detect. There have already been some efforts to study ZDMs in such contexts, for example in Ref.~\cite{Kokkotas:2015uma,Boyce:2025fpr,Cano:2025mht}, and they often involve introducing new fields coupled to the metric perturbations. This renders the calculation of quasinormal frequencies challenging, making \DF-like approximations a reasonable first step for analyzing ZDMs in such cases.

\section*{Acknowledgments}
We thank Aaron~Zimmerman for discussions, and Pablo A. Cano,
\'{O}scar~J.~C.~Dias, Nicola Franchini, Shahar~Hod, Jorge~E.~Santos, and
Alexander Zhidenko for correspondence.
We also thank Emanuele Berti, Fredric Hancock, Helvi Witek, and Kent Yagi for
their helpful feedback on this paper.
\appendix
\begingroup
\allowdisplaybreaks
\section{The real-valued potentials $V_s$} \label{app:transformed_potentials}
Here we present the \KN potentials $V_s$ appearing in Eq.~\eqref{eq:scat}, obtained after appropriate modifications to the analogous Kerr potentials given in Refs.~\cite{Yang:2013uba} and \cite{Detweiler:1977gy}.
For scalar perturbations ($s = 0$) we find:
\begin{equation}\label{eq:transformedpot_scalar}
    V_0(r) = -\frac{(K^2 -\lambda \Delta)}{(r^2+a^2)^2} + G^2 + \frac{\dd G}{\dd r_*}
\end{equation}
For electromagnetic perturbations ($s = -1$) we obtain,
\begin{align}\label{eq:transformedpot_em}
V_{-1} &= \frac{-K^2 + \lambda \Delta}{(r^2 + a^2)^2}
  - \frac{\Delta r(\Delta r + 2a^2)}{(r^2 + a^2)^4} \nonumber \\
  & - \frac{\Delta \left[\Delta(10r^2 + 2\nu^2)
  - (r^2 + \nu^2)(11r^2 - 5r + \nu^2)\right]}%
  {(r^2 + a^2)^2 \left[(r^2 + \nu^2)^2 + \eta \Delta\right]}\nonumber  \\
  & + \frac{6\Delta r (r^2 + \nu^2)^2\left[2\Delta r -(r^2 + \nu^2)(2r - 1) \right]}%
  {(r^2 + a^2)^2 \left[(r^2 + \nu^2)^2 + \eta \Delta\right]^2}\nonumber  \\
  & - \frac{\Delta(2r - 1)^2 \eta \left[2(r^2 + \nu^2)^2 - \eta \Delta\right]}%
  {4(r^2 + a^2)^2 \left[(r^2 + \nu^2)^2 + \eta \Delta\right]^2},
\end{align}
where we defined:
\begin{equation}\label{eq:a3}
        \nu^2 = a^2 - am/\omega,
        \quad \textrm{and} \quad
        \eta = (\kappa_{-1} - 2\lambda)/(4\omega^2).
\end{equation}
At last, for gravitational perturbations ($s = -2$), the potential reads:
\begin{equation}\label{eq:transformedpot_grav}
    V_{-2} = \frac{-K^2 + \lambda \Delta}{(r^2 + a^2)^2} + \frac{\Delta(b_2 p'\Delta)'}{(r^2+a^2)^2 b_2 p} + G^2 + \frac{\dd G}{\dd r_*},
\end{equation}
where a prime denotes a derivative with respect to $r$,
and where we defined:
\begin{equation}\label{eq:a5}
\begin{split}
    p &= (a_1 \Delta^2 + |\kappa_{-2}|)^{-1/2}, \\
    b_2 &= -\frac{8K^3}{\Delta^2} - \frac{4K}{\Delta}\left[\frac{(1 - 4a^2)}{2\Delta} - \lambda\right]
    -\frac{4\omega}{\Delta}(r-2a^2).
\end{split}
\end{equation}
\endgroup

\begin{figure}[b]
\includegraphics{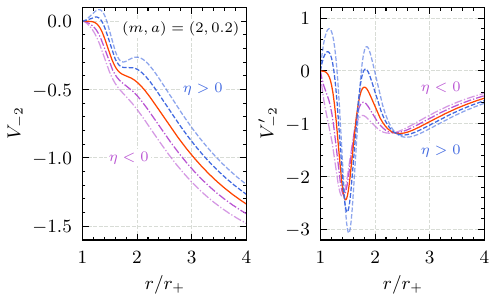}
\caption{The potential $V_{-2}$ (left panel) and its radial derivative $V_{-2}'$ (right panel) for $\eta \in [-1, 1]$, in increments of $0.5$, for $m = 2$ and black-hole spin $a=0.2$, as functions of $r$. The dashed and dot-dashed curves correspond to positive and negative values of $\eta$, respectively, and the solid curve corresponds to $\eta = 0$. For $\eta > 0$, corresponding to $V_{-2}''(r_+) > 0$, the potential $V_{-2}$ has peaks outside the horizon, indicating the presence of DMs alongside the ZDMs. For $\eta < 0$, no such peaks exist outside the horizon, since the potential is monotonically decreasing for $r > r_+$. Thus, in this case there are no peaks outside the horizon capable of supporting DMs. Similar results are obtained for other values of $m$ and $a$, as well as for other values of the spin of the perturbing field $s$.}
\label{fig:v_vdv}
\end{figure}
In Eqs.~\eqref{eq:a3} and \eqref{eq:a5}, the quantities $\kappa_{-1}$, $a_1$, and $\kappa_{-2}$ are given by Eqs.~(C4), (C9), and (C10), respectively, in Ref.~\cite{Yang:2013uba}.

\section{Extrema of the potential $V_s$} \label{app:potential_extrema}

In Sec.~\ref{sec:dsqfsq}, we claimed, without justification, that the sign of $V_s''(r_+)$ determines whether or not there exists a peak of $V_s(r)$ \textit{outside} the horizon [cf.~Eq.~\eqref{eq:d2V_dr2}].
We now justify this claim. First, we note that
\begin{equation}\label{eq:vs_inf}
    \lim_{r \rightarrow \infty} V_s(r) = -m^2,
\end{equation}
where $V_s$ are the potentials presented in Appendix~\ref{app:transformed_potentials}.
Combining Eq.~\eqref{eq:vs_inf} with Eq.~\eqref{eq:vs_hor_conds}, two distinct possibilities arise: if $V_s''(r_+) > 0$, then continuity automatically implies that $V_s(r)$ must have \textit{at least} one local maximum outside the horizon capable of supporting DMs. However, if $V_s''(r_+) < 0$, then the lack of any peaks outside the horizon is only guaranteed if $V_s(r)$ is \textit{monotonically decreasing} for $r>r_+$:
    \begin{equation}\label{eq:vs_monotonicity}
        V_s'(r) < 0, \quad r \in (r_+,\infty).
    \end{equation}
Given the complicated forms of the potentials $V_s(r)$, Eq.~\eqref{eq:vs_monotonicity} can only be verified numerically. We do so by defining an auxiliary quantity motivated by the form of Eq.~\eqref{eq:fsq_defn_kn}:
    \begin{equation}\label{eq:a_eta}
        A = g_s(a,m) + \eta, \quad \eta \in \mathbb{R}.
    \end{equation}
Here $A$ is the angular separation constant evaluated at $\omega = m\Omega_H |_{\rm ext}$, and $g_s(a,m)$ is a shorthand notation for the functions appearing in Eq.~\eqref{eq:fsq_defn_kn}.
Equation~\eqref{eq:a_eta} then implies that the conditions $\mathcal{F}_s^2>0$ and $\mathcal{F}_s^2<0$ (equivalently, $V_s''(r_+)<0$ and $V_s''(r_+)>0$) are equivalent to $\eta<0$ and $\eta>0$, respectively.

In Fig.~\ref{fig:v_vdv}, we show $V_{-2}(r)$ and $V_{-2}'(r)$, for $\eta \in [-1, 1]$ in steps of $0.5$, for $m=2$ and $a=0.2$.
For $\eta > 0$, $V_{-2}(r)$ has peaks outside the horizon, thereby indicating the presence of DMs alongside the ZDMs.
For $\eta < 0$, there are no peaks outside the horizon, since the potential is monotonically decreasing for $r > r_+$; thus, DMs are absent in this case.
Similar results hold for other values of $(m,a)$, and for different spins $s$ of the perturbing field.

\bibliography{biblio}

\end{document}